\documentclass[twocolumn]{aastex631}

\usepackage{color}
\usepackage{amsmath}

\newcommand\mr{\mathrm}

\begin{document}
\title{Stellar Tidal Disruptions by Newborn Neutron Stars or Black Holes: A Mechanism for Hydrogen-poor (Super)luminous Supernovae and Fast Blue Optical Transients}
\author[0000-0002-6347-3089]{Daichi Tsuna}
\affiliation{TAPIR, Mailcode 350-17, California Institute of Technology, Pasadena, CA 91125, USA}
\affiliation{Research Center for the Early Universe (RESCEU), School of Science, The University of Tokyo,  Bunkyo-ku, Tokyo 113-0033, Japan}
\author[0000-0002-1568-7461]{Wenbin Lu}
\affiliation{Department of Astronomy and Theoretical Astrophysics Center, University of California at Berkeley, Berkeley, CA 94720, USA}

\correspondingauthor{Daichi Tsuna}
\email{tsuna@caltech.edu}

\begin{abstract}
Hydrogen-poor supernovae (SNe) of Type Ibc are explosions of massive stars that lost their hydrogen envelopes, typically due to interactions with a binary companion. We consider the case where the natal kick imparted to the neutron star (NS) or black hole (BH) remnant brings the compact object to a collision with a main-sequence companion, eventually leading to full tidal disruption of the companion.
Subsequently, super-Eddington accretion onto the NS/BH launches a powerful, fast wind which collides with the SN ejecta and efficiently converts the kinetic energy of the wind into radiation. The radiation is reprocessed by the surrounding ejecta into a luminous ($\sim 10^{44}$ erg s$^{-1}$ at peak), days to months-long transient with optical peaks from $-19$ to $-21$ mag, comparable to (super)luminous Type Ibc SNe and fast blue optical transients (FBOTs) like AT2018cow. From a Monte-Carlo analysis we estimate the fraction of tidal disruptions following SNe in binaries to be $\sim 0.1$--$1$\%, roughly compatible with the event rates of these luminous SNe. At the broad-brush level, our model reproduces the multi-wavelength and spectral observations of FBOTs, and has the potential to explain peculiar features seen in some (super)luminous SNe which are difficult to reproduce by the conventional magnetar spindown mechanism, such as late-time hydrogen lines, bumpy light curves, and pre-peak excess. 
\end{abstract}

\section{Introduction}
Core-collapse supernovae (SNe) have a large diversity in their photometric and spectroscopic appearances, which reflects the diversity in the progenitor's evolution and mass loss \citep[e.g.][]{Smartt09,Smith14}. The spectroscopic signature divides core-collapse SNe into two main types: Type II SNe from stars having hydrogen-rich envelopes, and Type Ibc SNe from stars that lost their hydrogen-rich envelope well before core-collapse.

The leading channel for Type Ibc SNe is massive stars whose hydrogen-rich envelopes had been stripped off by binary interaction, typically with a main-sequence companion \citep[e.g.,][]{Shigeyama1990,Podsiadlowski92,Eldridge08,Smith11}. This has been supported by the high event rates ($\sim 30\%$ of core-collapse SN; \citealt{Smith11}), low inferred ejecta masses \citep[e.g.,][]{Drout11,Lyman16,Taddia18} and lack of detections of high-mass progenitors \citep[][]{Eldridge13,Smartt15}, all of which disfavor very massive single stars (with initial masses $\gtrsim 25$--$30~M_\odot$) as the dominant channel.

When the stripped star undergoes core-collapse and explodes, its remnant, typically a neutron star (NS), receives a natal kick \citep{Lyne94,Hobbs05} due to asymmetry in the SN explosion. Recent simulations showed that the collapse of high-compactness stellar cores can sometimes also lead to black hole (BH) formation, together with strong, asymmetric explosions that give rise to large kicks to the remnant \citep{burrows23_SN_from_BH, Burrows24}. The most common outcomes of the binary are that they would either survive or be unbound, depending on the direction and magnitude of the kick. However, in rare cases where the kick is sufficiently large and directed to the companion, there is a third possibility: collision of the NS/BH and the companion \citep[e.g.,][]{Perets16,Hirai22}. 

Modeling of close encounters between a compact object and a main-sequence star have predicted various phenomenological outcomes \citep[e.g.,][]{Davies92,Lombardi06,Perets16,Wang21,Kremer22_TDE,Kremer22_MSP,Hirai22,Kremer23,Vynatheya24,Everson24,Hutchinson-Smith24,Kiroglu25a,Kiroglu25b}. An important finding is that for sufficiently close encounters, the star can be tidally disrupted and leave part of it bound to the compact object. The bound debris would circularize and accrete onto the compact object, typically at highly super-Eddington rates. 

\begin{figure*}[]
    \centering
    \includegraphics[width=0.9\linewidth]{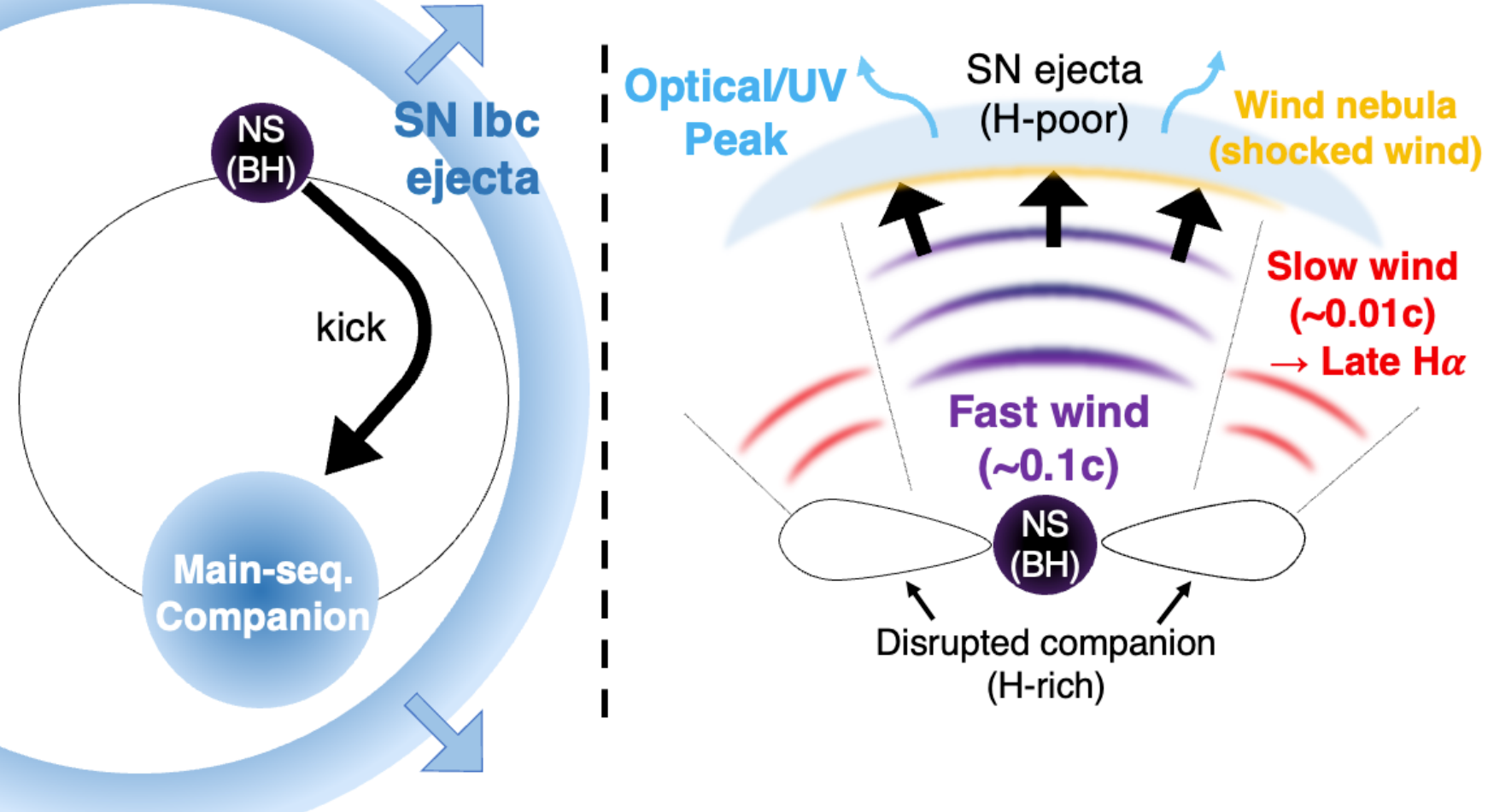}
    \caption{Schematic picture of our model for Type I (super-)luminous SNe and fast blue optical transients (not to scale). A newborn compact object from a stripped-envelope (Type Ibc) SN receives a natal kick, that leads to encounter with its main-sequence companion. The subsequent disruption and circularization of the companion result in strong outflows via super-Eddington accretion onto the compact object, energizing the SN to luminosities of $\sim 10^{44}\ {\rm erg\ s^{-1}}$. The slow wind ($\sim 0.01c$) embedded in the SN ejecta, launched from the outer part of the disk, can contribute to late-time H$\alpha$ emission after the ejecta becomes transparent.
    }
    \label{fig:schematic}
\end{figure*}

In this work we explore the observational signatures of such events, building a model for the transient including the energy injection due to super-Eddington accretion onto the newborn NS/BH. We find that the strong outflow can energize the SN ejecta to power transients with luminosities of the order of $10^{44}$ erg s$^{-1}$ for days to months. Such luminosities are comparable to Type I luminous and superluminous SNe (SLSNe) \citep{Moriya18,Gal-Yam19,Nicholl21,Gomez22,Gomez24} and fast blue optical transients (FBOTs) like AT2018cow \citep[e.g.,][]{Prentice18,Perley19, Ho19,Margutti19}, which are 10-100 times more luminous at peak than typical Type Ibc SNe. 

The origins for both SLSNe \citep[e.g.,][]{Kasen10,Woosley10,Chevalier11,Dexter13,Metzger15,Soker17} and FBOTs \citep[e.g.,][]{Margutti19,Lyutikov19,Quataert19,Soker19,Leung20,Uno20,Kremer21,Metzger22} are under debate. The brightness of these transients likely require energy injection by a central compact object, but the nature of this central engine is an open question. While the popularly invoked hypothesis is based on a rapidly spinning magnetar \citep{Kasen10,Woosley10}, a significant fraction of SLSNe have peculiar properties that the magnetar model does not naturally explain, such as late-time hydrogen lines \citep[][also seen in FBOTs; \citealt{Perley19,Margutti19,Gutierrez24}]{Yan15,Yan17} and light curve bumps \citep[e.g,][]{Nicholl16,Inserra17,Hosseinzadeh22}. We show that our model can explain the overall observations of these luminous transients, including the spectral and photometric complexities seen in many of these objects.

This paper is organized as follows. In Section \ref{sec:model} we explain the physical ingredients of our model, whose properties are schematically summarized in Figure \ref{fig:schematic}. We show the resulting light curve and their properties (peak magnitude, rise time) in Section \ref{sec:results}. In Section \ref{sec:discussion} we discuss the expected event rates of these events, as well as potential connections of our model to the aforementioned peculiarities in SLSNe and to the multi-wavelength observations of FBOTs like AT2018cow. We conclude in Section \ref{sec:conclusion}.

\section{Model}
\label{sec:model}
We consider a binary system composed of a stripped SN progenitor with a main-sequence companion. Such systems are common channels for stripped envelope SNe of Type Ibc, as massive stars mostly exist in binaries and $\sim 1/3$ of the primary is stripped by mass transfer onto the companion \citep{Sana12}. The binary separation at core-collapse of the primary is expected to be $a_{\rm bin}\sim 10^{12}$--$10^{13}\ {\rm cm}$ \citep{Moriya15}, whose uncertainties depend on prescriptions for mass transfer and angular momentum loss during envelope stripping. Assuming a circular binary orbit, the pre-SN orbital velocity is 
\begin{eqnarray}
    v_{\rm orb}&=&\sqrt{\frac{G(M_{\rm prog}+M_*)}{a_{\rm bin}}} \nonumber \\
    &\sim& 120\ {\rm km \ s^{-1}}\left(\frac{M_{\rm prog}+M_*}{10~M_\odot}\right)^{1/2}\left(\frac{a_{\rm bin}}{10^{13}\ {\rm cm}}\right)^{-1/2},
\end{eqnarray}
where $M_{\rm prog}$ is the mass of the (stripped) SN progenitor, $M_*$ is the mass of the companion, and $G$ is the gravitational constant. The companion is expected to be a main-sequence star for most cases, whose mass has a distribution peaking at $5$--$10~M_\odot$ depending on metallicity and assumptions on parameters regarding mass transfer \citep[][]{Zapartas17}. 

The stripped star undergoes a core-collapse SN, with an ejecta mass $M_{\rm ej}$ and explosion energy $E_{\rm exp}$. The SN ejecta has a bulk velocity
\begin{eqnarray}
    v_{\rm ej,0}&\sim &\sqrt{2E_{\rm exp}/M_{\rm ej}} \nonumber \\
    &\approx & 6000\ {\rm km\ s^{-1}}\left(\frac{E_{\rm exp}}{10^{51}\ {\rm erg}}\right)^{1/2}\left(\frac{M_{\rm ej}}{3\ M_\odot}\right)^{-1/2} \gg v_{\rm orb}. \nonumber
\end{eqnarray}
In our fiducial case, we assume that the explosion leaves a NS remnant of mass $M_{\rm NS}=M_{\rm prog}-M_{\rm ej}=1.4~M_\odot$, and the case of a BH remnant will be discussed in Section \ref{sec:disk_BH}. As the ejecta immediately leaves the binary, the remnant NS receives an instantaneous kick due to asymmetry in the SN explosion. If the magnitude of the kick $v_{\rm kick}$ is comparable to or larger than $v_{\rm orb}$, there is a small chance that the kick is oriented at a direction which brings the NS to encounter the star. When $v_{\rm kick}\gg v_{\rm orb}$, the kicked NS encounters (shoots into) the star at a time after the SN explosion of
\begin{eqnarray}
    t_{\rm enc}\sim \frac{a_{\rm bin}}{v_{\rm kick}} \sim 3\ {\rm day} \left(\frac{a_{\rm bin}}{10^{13}~{\rm cm}}\right) \left(\frac{v_{\rm kick}}{400~{\rm km\ s^{-1}}}\right)^{-1},
\end{eqnarray}
while if $v_{\rm kick}$ is comparable to $v_{\rm orb}$, they encounter at a time of roughly half the initial orbital period
\begin{eqnarray}
    t_{\rm enc}&\sim& \frac{\pi a_{\rm bin}}{v_{\rm orb}} \nonumber\\
    &\sim& 30\ {\rm day} \left(\frac{a_{\rm bin}}{10^{13}\ {\rm cm}}\right)^{3/2}\left(\frac{M_{\rm prog}+M_*}{10~M_\odot}\right)^{-1/2}.
\end{eqnarray}

We focus on the case where the encounter leads to tidal disruption of the star, as obtained from recent smoothed particle hydrodynamics simulations of star-compact object encounters \citep{Kremer22_TDE,Kremer23}. The bound part of the debris is expected to form a thick rotating disk around the remnant. The accretion rate of the disk is orders of magnitude larger than the Eddington rate \citep{Kremer22_TDE}, and we expect radiation-driven winds due to super-Eddington accretion. The fast part of the disk wind can interact with the outer SN ejecta, which can (re-)energize the SN ejecta and power a luminous transient. We hereafter model the detailed transient emission expected for these systems.

\subsection{Formation and evolution of the accretion disk}
An important point regarding the tidal effects on such encounters is that as $M_*$ is typically larger than $M_{\rm NS}$, the conventionally defined tidal radius $r_{\rm T} = (M_{\rm NS}/M_*)^{1/3}R_*$ is inside the star, where $R_*$ is the stellar radius. The dynamics of the disruption and subsequent disk formation would thus be very different from tidal disruption events (TDEs) by supermassive BHs, where $r_{\rm T}\sim (M_{\rm BH}/M_*)^{1/3}R_*\gg R_*$ since the BH mass $M_{\rm BH}$ is much greater than $M_*$.

\cite{Kremer22_TDE} simulated the hydrodynamics of the collision of a main-sequence star with a BH companion including mass ratios of $M_{\rm BH}/M_*<1$, and found full disruptions for encounters with pericenter distances of (see the dashed line in their Figure 1)
\begin{eqnarray}
    r_{\rm p}\lesssim R_* \sim 3\times 10^{11}\ {\rm cm} \left(\frac{M_*}{10~M_\odot}\right)^{0.6},
\end{eqnarray}
which includes encounters that are fully disrupted upon first passage ($r_{\rm p}\lesssim 0.4r_{\rm T} \sim 6\times 10^{10}{\rm cm} (M_*/10~M_\odot)^{4/15}$), and those that are bound with short orbital periods by tidal capture and become disrupted within several additional encounters. We hereafter adopt the mass-radius relation $R_*\approx R_\odot(M_*/M_\odot)^{0.6}$ for main-sequence stars of $M_*\gtrsim 1~M_\odot$ \citep{Kippenhahn}. Such encounters occur for a small fraction of the entire SN Ibc events, whose rates are estimated in Section \ref{sec:rates}.

The time of full stellar disruption $t_{\rm TDE}$ depends on the post-SN pericenter radii. For very deep penetrations ($r_{\rm p}\lesssim 0.4r_{\rm T}$; \citealt{Kremer22_TDE}) we expect full disruption in the first passage, likely indicating $t_{\rm TDE}\approx t_{\rm enc}$. For shallower penetrations of $r_{\rm p}\approx r_{\rm T}$, the star can survive for several orbits before it is eventually disrupted, as seen in selected long-term simulations of \cite{Kremer22_TDE} (their Figure 9; see also \citealt{Kremer23}). In this work we remain agnostic to the history of the NS-star binary before the full disruption, and focus on the full disruption where a dominant fraction of the original star is disrupted. We discuss the possible effects of multiple passages in Section \ref{sec:multiple_encounter}.

The simulations by \cite{Kremer22_TDE} found that in the case of full disruption of a massive star with $1\leq M_*/M_{\rm BH}\leq 2$, a majority ($60$--$80$\%) of the star is bound to the compact object, which returns and forms a rotationally supported disk with characteristic radius of roughly $\sim 2R_*$ (their Table 2). However, there are two limitations when applying this to our case where the mass ratio between the star and the compact object $M_*/M_{\rm NS}$ is likely even larger. First, the NS is expected to gravitationally capture only material comparable to its own mass, and the rest of the star that is not gravitationally bound to the NS would likely be ejected by the strong feedback from the accretion-driven wind. Second, due to limited angular momentum budget (roughly the orbital angular momentum of the NS), the characteristic disk radius that governs the timescale for viscous accretion (eq. \ref{eq:t_visc}) may be smaller than $2R_*$ and closer to the Bondi radius of the NS, roughly $2GM_{\rm NS}/(GM_*/R_*)\sim (2M_{\rm NS}/M_*)R_*$. We thus take the fiducial values of the initial disk parameters as
\begin{eqnarray}
    M_{\rm disk,0} &=& M_{\rm NS} \label{eq:Mdisk_0}\\
    r_{\rm disk,0} &=& \left(\frac{2M_{\rm NS}}{M_*}\right)R_* \sim 8\times 10^{10}{\rm cm}\left(\frac{M_*}{10~M_\odot}\right)^{-0.4}. 
 \label{eq:Rdisk_0}
\end{eqnarray}

For a disk of mass $M_{\rm disk,0}$ at characteristic radius of $r_{\rm disk,0}$, the initial viscous time is given as
\begin{eqnarray}
    t_{\rm visc,0}&\approx& \frac{1}{\alpha (H/R)^2}\sqrt{\frac{r_{\rm disk,0}^3}{G(M_{\rm disk,0}+M_{\rm NS})}} \nonumber \\
    &\sim& 1.4\ {\rm day}\left(\frac{M_*}{10~M_\odot}\right)^{-0.6} \left(\frac{\alpha}{0.1}\right)^{-1} \left(\frac{H/R}{0.3}\right)^{-2},
    \label{eq:t_visc}
\end{eqnarray}
where $\alpha$ is the viscosity parameter, $H/R$ is the height to radius ratio of the disk. The accretion rate over this timescale is $\sim M_{\rm disk,0}/t_{\rm visc,0}\sim 400~M_\odot\ {\rm yr}^{-1}$, which is $\sim 10$ orders of magnitude larger than the Eddington accretion rate for NSs.
The disk spreads over time to larger radii due to angular momentum transport, and is also prone to mass loss via outflows as the high optical depth of inflowing matter leads to inefficient radiative cooling \citep{Narayan95,Blandford99}. 

We construct a one-zone model for the evolution of the radius $r_{\rm disk}(t)$ and mass $M_{\rm disk}(t)$ of the disk, with initial conditions in equations (\ref{eq:Mdisk_0}) and (\ref{eq:Rdisk_0}). For a disk with Keplerian rotation, its angular momentum is 
\begin{eqnarray}
    J_{\rm disk}&\approx& \int^{M_{\rm disk}}_{0}[G(M_{\rm NS}+M)r_{\rm disk}]^{1/2} dM \label{eq:Jdisk} \\
    &=& \frac{2}{3}(GM_{\rm NS}^3r_{\rm disk})^{1/2}\left[-1+\left(1+\frac{M_{\rm disk}}{M_{\rm NS}}\right)^{\!\!3/2}\right],
\end{eqnarray}
where we take care that $M_{\rm disk}$ is comparable to $M_{\rm NS}$ in contrast to the case of TDEs by supermassive BHs. We numerically integrate with time the following equations
\begin{eqnarray}
    \frac{dM_{\rm disk}}{dt} &=& -M_{\rm disk}/t_{\rm visc}, \label{eq:Mdot}\\
    \frac{d{J}_{\rm disk}}{dt} &=& -F_{\rm w}\frac{J_{\rm disk}}{t_{\rm visc}}, \label{eq:Jdot} \\
    t_{\rm visc} &\approx& \frac{\sqrt{r_{\rm disk}^3/GM_{\rm NS}}}{\alpha (H/R)^2}\frac{2M_{\rm NS}}{M_{\rm disk}}\left[\sqrt{1+\frac{M_{\rm disk}}{M_{\rm NS}}} -1\right], \label{eq:avg_tvis} 
\end{eqnarray}
where $F_{\rm w}$ is the ratio of the wind's specific angular momentum to that of the disk \citep{Shen14}. As typically modeled for such accretion flows, we prescribe the mass inflow rate as a power-law in radius from the NS as $\dot{M}(r)\propto r^{p}$ \citep{Blandford99}, valid from $r_{\rm disk}$ to the NS radius $R_{\rm NS}\approx 12$ km. A range of $0.3\leq p \leq 0.8$ is suggested from numerical simulations of advection-dominated flows \citep{Yuan&Narayan2014}, and we adopt $p=0.5$ in this work as motivated by more recent simulations \citep{Cho24_ADAF_accretion_profile, guo24_ADAF_accretion_profile}.

Because $R_{\rm NS}\ll r_{\rm disk}$, the mass loss in eq. (\ref{eq:Mdot}) is almost entirely due to the disk outflow instead of accretion onto the NS. We take the angular momentum of the wind at each radius from the NS to be equivalent to that of the disk, which results in $F_{\rm w}=2p/(2p+1)$, i.e. $F_{\rm w}=0.5$ for $p=0.5$ \citep{Shen14,Kremer19}. We adopt a viscous time mass-averaged over the disk similar to the treatment in eq. (\ref{eq:Jdisk})\footnote{For $M_{\rm disk}\ll M_{\rm NS}$ we recover $J_{\rm disk}=(GM_{\rm NS}r_{\rm disk})^{1/2}M_{\rm disk}$ and $t_{\rm visc}=\alpha^{-1}(H/R)^{-2}\sqrt{r_{\rm disk}^3/GM_{\rm NS}}$, the equations commonly seen in one-zone disk models in this regime.}, but the uncertainties there (expected to be of order unity) can be absorbed into the parameter $\alpha(H/R)^2$.

\begin{figure}[]
    \centering
    \includegraphics[width=\linewidth]{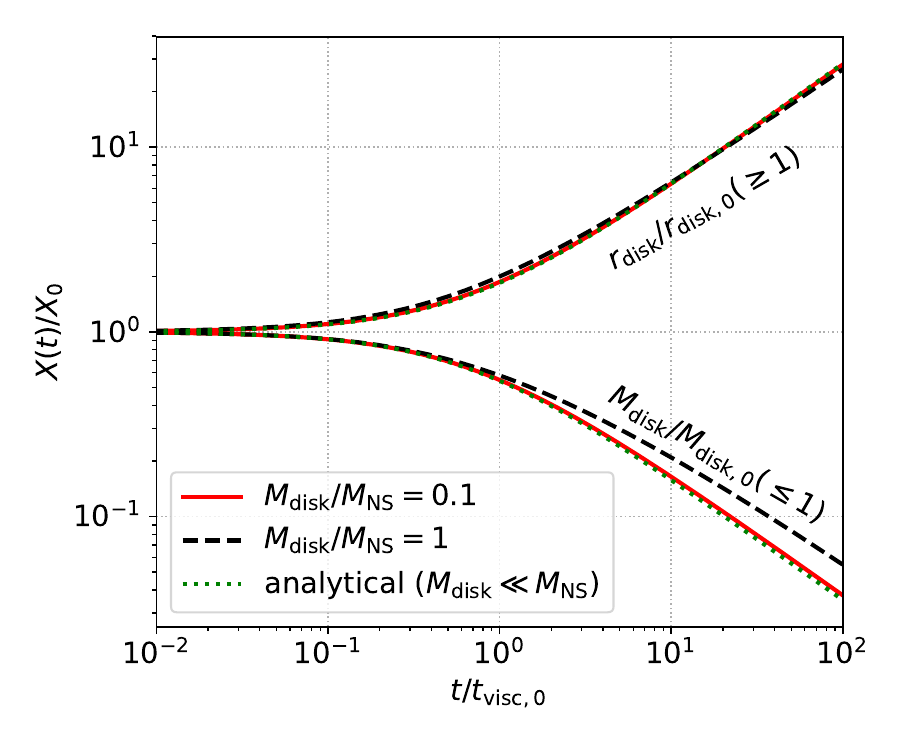}
    \caption{Evolution of the disk mass and radius, for two cases of $M_{\rm disk}$ with $M_{\rm disk}/M_{\rm NS}=0.1$ and $M_{\rm disk}/M_{\rm NS}=1$. Dotted lines (almost overlapping with the solid lines) show the analytical solution under $M_{\rm disk}\ll M_{\rm NS}$. We assume $r_{\rm disk,0}=R_\odot$, $M_{\rm NS}=1.4M_\odot$ for the numerical calculations.
    }
    \label{fig:disk_evolve}
\end{figure}

Figure \ref{fig:disk_evolve} shows the obtained evolutions of the disk mass and radius, for two cases of of $M_{\rm disk,0}/M_{\rm NS}=0.1, 1$. The former may be achieved for partial disruption of the star, whereas the latter is a more realistic value for full disruptions. For the case of $M_{\rm disk,0}/M_{\rm NS}=0.1$, we recover the analytical solution for $M_{\rm disk,0}\ll M_{\rm NS}$ of 
\begin{eqnarray}
    r_{\rm disk}/r_{\rm disk,0} &=& [1+(3-3F_{\rm w})t/t_{\rm visc,0}]^{2\over 3} \nonumber\\
    M_{\rm disk}/M_{\rm disk,0} &=& [1+ (3-3F_{\rm w})t/t_{\rm visc,0}]^{-{1\over 3(1-F_{\rm w})}}.
\end{eqnarray}
For the case with $M_{\rm disk,0}\sim M_{\rm NS}$, we find the evolution of these to be slightly slower than the above solutions.

We assume that the wind launched at a radius $r$ from the NS has a positive specific energy of $GM_{\rm NS}/2r$, with asymptotic velocity $v_{\rm wind}(r)\approx\sqrt{GM_{\rm NS}/r}$. The kinetic luminosity of the wind at a given time $t$ is then obtained by integrating over $r$ as
\begin{eqnarray}
    L_{\rm kin}(t)&\approx &\frac{p}{2(1-p)}\frac{GM_{\rm NS}|\dot{M}_{\rm disk}|}{R_{\rm NS}} \left(\frac{R_{\rm NS}}{r_{\rm disk}}\right)^{p},
\end{eqnarray}
where we have used $R_{\rm NS}\ll r_{\rm disk}$. 

From the analysis above, we can obtain an order-of-magnitude estimate of the energy budget of the wind. From Figure \ref{fig:disk_evolve}, a fraction of $\eta_{\rm w}\approx 50\%$ of the initial disk mass is lost via wind within the first $t_{\rm visc,0}$. Adopting $p=0.5$, the average kinetic luminosity and total energy of the wind over the first $t_{\rm visc,0}$ are roughly
\begin{eqnarray}
    L_{\rm kin,0}&\sim & \frac{1}{2}\frac{GM_{\rm NS}}{R_{\rm NS}} \frac{\eta_{\rm w}M_{\rm disk,0}}{t_{\rm visc,0}}\left(\frac{R_{\rm NS}}{r_{\rm disk,0}}\right)^{0.5} \nonumber \\
    &\sim& 3\times 10^{45}~{\rm erg\ s^{-1}} \left(\frac{M_*}{10~M_\odot}\right)^{0.8} \left(\frac{\alpha}{0.1}\right)\left(\frac{H/R}{0.3}\right)^{2} \\
    E_{\rm kin,0} &\sim& L_{\rm kin,0}t_{\rm visc,0} \sim 4\times 10^{50}\ {\rm erg}\left(\frac{M_*}{10~M_\odot}\right)^{0.2}, \label{eq:E_wind}
\end{eqnarray}
where $|\dot{M}_{\rm disk}|\approx \eta_{\rm w}M_{\rm disk,0}/t_{\rm visc,0}$. The fastest part of the wind, carrying the dominant fraction of the energy, can catch up with the SN ejecta and inject its energy via shocks. If this energy can be efficiently converted to radiation, this can make the SN much brighter than normal core-collapse SNe.

\subsection{Wind Energy Injection into the SN ejecta}
\label{sec:nebula}
The shocked region between the fast disk wind and the SN ejecta forms a ``wind nebula", where energy is injected from the disk wind to the ejecta. We solve the radial propagation of the wind nebula $R_{\rm neb}(t)$ under a thin-shell approximation, largely following the semi-analytical model constructed for SNe powered by a magnetar wind \citep[][Appendix B]{Kashiyama16}.

We adopt a power-law density structure of the initial (unshocked) SN ejecta under homologous expansion
\begin{eqnarray}
\rho_{\rm ej}(R)=\frac{3-\delta}{4\pi}\frac{M_{\rm ej}}{R_{\rm ej}^3}\left(\frac{R}{R_{\rm ej}}\right)^{-\delta}\ (R\leq R_{\rm ej}),
\label{eq:rho_ej}
\end{eqnarray}
where we set $\delta=1$ \citep{Chevalier89}. 
At a given time, only the disk wind launched from within a critical radius $r_{\rm crit}$ from the NS, with an asymptotic velocity greater than $dR_{\rm neb}/dt$, can contribute to energy injection. Then, the disk wind has an injection luminosity and mass outflow rate of
\begin{eqnarray}\label{eq:Lwind_Mdotwind_injection}
     L_{\rm wind} &=&\frac{p}{2(1-p)}\frac{GM_{\rm NS}|\dot{M}_{\rm disk}|}{r_{\rm disk}}\left[\left(\frac{R_{\rm NS}}{r_{\rm disk}}\right)^{p-1} - \left(\frac{r_{\rm crit}}{r_{\rm disk}}\right)^{p-1} \right] \label{eq:Lwind}\\
     \dot{M}_{\rm wind} &=& |\dot{M}_{\rm disk}| \left[\left(\frac{r_{\rm crit}}{r_{\rm disk}}\right)^{p} - \left(\frac{R_{\rm NS}}{r_{\rm disk}}\right)^{p} \right],
\end{eqnarray}
where $r_{\rm crit}$ is a critical radius below which the wind is sufficiently fast to catch up with the wind nebula 
\begin{eqnarray}
    r_{\rm crit}(t) \approx {\rm min}\left[\frac{GM_{\rm NS}}{(dR_{\rm neb}/dt)^2}, r_{\rm disk}(t)\right].
    \label{eq:r_crit}
\end{eqnarray}
The injection luminosity is dominated by the fast wind launched near the NS surface, whereas the mass outflow rate is dominated by the slow wind launched near $r_{\rm crit}$ (see also Appendix \ref{sec:radiative power}). However, the ram pressure of the wind ($\propto \dot{M}(r)v_{\rm wind}(r)$), which sets the nebula's expansion, has roughly equal contribution from all radii from the NS for our adopted value of $p=0.5$. Note that only a fraction of the wind-injected luminosity $L_{\rm wind}$ gets converted into radiation by free-free emission and inverse-Compton scattering in the shocked wind region. The radiative efficiency $\epsilon_{\rm rad}$ of the shocked wind region is incorporated into our light curve model in Section \ref{sec:onezone_model}, and the detailed calculation of $\epsilon_{\rm rad}$ (eq. \ref{eq:wind_radiative_efficiency_global}) will be presented in the Appendix \ref{sec:radiative power}.

We evolve $R_{\rm neb}$ following \cite{Kashiyama16} as
\begin{eqnarray}
    \frac{dR_{\rm neb}}{dt} = \tilde{v}_{\rm neb} + \frac{R_{\rm neb}}{t}
\end{eqnarray}
where $\tilde{v}_{\rm neb}$ is the velocity of the shocked wind in the rest frame of the unshocked ejecta. We solve for $\tilde{v}_{\rm neb}$ from pressure equilibrium in the comoving frame of the shocked wind,
\begin{eqnarray}
    &&\rho_{\rm ej}\tilde{v}_{\rm neb}^2 = \rho_{\rm wind}\left[\langle v_{\rm wind} \rangle - \left(\tilde{v}_{\rm neb} + \frac{R_{\rm neb}}{t}\right) \right]^2 \nonumber \\
    \to&& \tilde{v}_{\rm neb} = \frac{1}{1+\sqrt{\rho_{\rm ej}/\rho_{\rm wind}}} \left( \langle v_{\rm wind} \rangle - \frac{R_{\rm neb}}{t} \right). \label{eq:vtilde_neb}
\end{eqnarray}
As the ram pressure is equally contributed from the entire wind, we here defined a characteristic wind velocity
\begin{eqnarray}
    \langle v_{\rm wind} \rangle &=& \sqrt{\frac{2L_{\rm wind}}{\dot{M}_{\rm wind}}} \nonumber \\
    &=& \sqrt{\frac{p}{1-p} \frac{GM_{\rm NS}}{R_{\rm NS}}\frac{1-\left(r_{\rm crit}/R_{\rm NS}\right)^{p-1}}{\left(r_{\rm crit}/R_{\rm NS}\right)^p - 1}},
    \label{eq:vwind_charact}
\end{eqnarray}
and a characteristic upstream wind density
\begin{eqnarray}
\rho_{\rm wind}(t)\approx\frac{\dot{M}_{\rm wind}(t)}{4\pi R^2_{\rm neb}(t) \langle v_{\rm wind}(t)\rangle},    
\end{eqnarray}
where we assume the wind travel time $R_{\rm neb}/\langle v_{\rm wind}\rangle\ll t$. We have also used $\tilde{v}_{\rm neb}>0, \langle v_{\rm wind} \rangle>(\tilde{v}_{\rm neb}+R_{\rm neb}/t)$ to obtain the sign for eq. (\ref{eq:vtilde_neb}). The initial conditions for $R_{\rm neb}$ and $\tilde{v}_{\rm neb}$ at $t=t_{\rm TDE}$ are set as $R_{\rm neb}=R_{\rm disk,0}, \tilde{v}_{\rm neb}=0$. When $R_{\rm neb}$ reaches $R_{\rm ej}$, we fix this to $R_{\rm neb}=R_{\rm ej}$.

\begin{figure*}
    \centering
    \includegraphics[width=\linewidth]{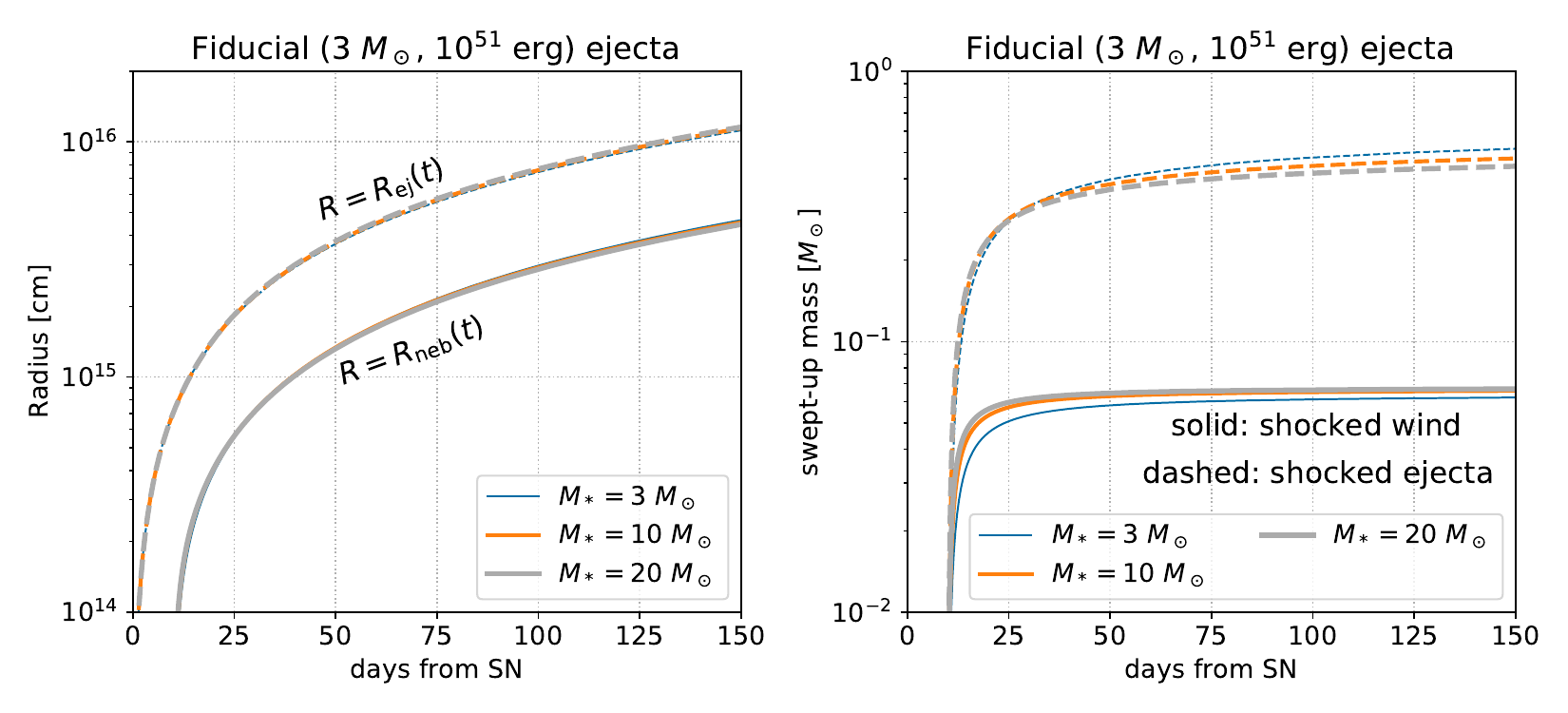}
    \caption{Dynamics of the wind nebula solved in Section \ref{sec:nebula} for a typical SN Ibc ejecta with $M_{\rm ej}=3~M_\odot, E_{\rm exp}=10^{51}$ erg. Left panel: Time evolution of the wind nebula radius $R_{\rm neb}(t)$ (solid lines) and the characteristic ejecta radius $R_{\rm ej}(t)$ (dashed lines), varying $M_*$. Note that the lines for different $M_*$ nearly overlap. Right panel: Time-dependent masses of the swept-up disk wind (solid lines) and the swept-up SN ejecta $M_{\rm ej}(r<R_{\rm neb}(t))$ (dashed lines), varying $M_*$. For both panels, other model parameters are fixed to the fiducial values in Table \ref{tab:params}.
    }
    \label{fig:nebula_prop}
\end{figure*}

Figure \ref{fig:nebula_prop} shows the time evolution of $R_{\rm neb}(t)$, the time-integrated mass of the shocked wind $M_{\rm w,sh}(t)=\int^t_{t_{\rm TDE}} \dot{M}_{\rm wind}(t')dt'$ and the mass of the shocked ejecta $M_{\rm ej, sh}(t)=M_{\rm ej}(r<R_{\rm neb}(t))$. We have chosen the fiducial model parameters in Table \ref{tab:params} as $t_{\rm TDE}=10$ day and $\alpha(H/R)^2=0.01$ for the disk evolution, and varied $M_*$. For lower $M_*$ both $R_{\rm neb}$ and the swept-up ejecta mass are slightly lower initially but higher at late times, due to the slower evolution of viscous accretion. However, the dependence on $M_*$ is very weak.

We overplot the characteristic ejecta radius $R_{\rm ej}$, which is evolved by a one-zone model in the next section. The wind nebula is generally embedded in the SN ejecta, with asymptotic velocities of $dR_{\rm neb}/dt\approx 4000$ km s$^{-1}$. This gives rough estimates on $r_{\rm crit}$ (eq. \ref{eq:r_crit}) of
\begin{eqnarray}
    r_{\rm crit}\sim 1\times 10^9\ {\rm cm}\left(\frac{dR_{\rm neb}/dt}{4000~{\rm km\ s^{-1}}}\right)^{-2},
    \label{eq:r_crit_OM}
\end{eqnarray}
and a characteristic wind velocity (eq. \ref{eq:vwind_charact}, for $p=0.5$)
\begin{eqnarray}
    \langle v_{\rm wind} \rangle \sim 0.07c \left(\frac{dR_{\rm neb}/dt}{4000~{\rm km\ s^{-1}}}\right)^{0.5},
\end{eqnarray}
where $c$ is the speed of light. As $r_{\rm crit}\ll r_{\rm disk}$, the mass of the wind impacting the ejecta is small ($M_{\rm w, sh}\lesssim 0.1~M_\odot$), and much smaller than the ejecta mass $M_{\rm ej}$.

\subsection{Light Curve Modeling}
\label{sec:onezone_model}
We calculate the light curves by solving the evolution of the SN ejecta under wind injection from the NS disk. We adopt a one-zone model for the thermodynamics of the expanding SN ejecta, taking into account heating by energy deposition and acceleration via PdV work \citep[e.g.,][]{Arnett80,Arnett82,Kasen10,Dexter13,Nicholl17,Omand24,Tsuna24}. We solve the following set of equations
\begin{eqnarray}
    \frac{dR_{\rm ej}}{dt} &=& v_{\rm ej} \label{eq:drdt}, \\
    \frac{dE_{\rm rad}}{dt} &=&-(2-f_{\rm rad})\frac{E_{\rm rad}}{R_{\rm ej}} v_{\rm ej} \nonumber\\
    &&+ \left(1-e^{-\tau_{\rm ej}}\right) \epsilon_{\rm rad} L_{\rm wind} + L_{\rm Ni} - L_{\rm rad} \label{eq:dEraddt}, \\
    \frac{dE_{\rm gas}}{dt} &=&-(2-f_{\rm rad})\frac{E_{\rm gas}}{R_{\rm ej}} v_{\rm ej} + (1-\epsilon_{\rm rad})L_{\rm wind}  \label{eq:dEgasdt}, \\
    \frac{dE_{\rm kin}}{dt} &=& (2-f_{\rm rad})\frac{(E_{\rm rad}+E_{\rm gas})v_{\rm ej}}{R_{\rm ej}} \label{eq:dEkindt}, \\
    L_{\rm rad} &=& \frac{E_{\rm rad}}{t_{\rm diff}}, \label{eq:L_rad}
\end{eqnarray}
where $E_{\rm rad}, E_{\rm gas}$ are the internal energy in the SN ejecta and the wind nebula carried by thermal radiation and gas respectively, $f_{\rm rad}= E_{\rm rad}/(E_{\rm rad} + E_{\rm gas})$ is the fraction of internal energy carried by radiation, and $E_{\rm kin}=[(3-\delta)/2(5-\delta)]M_{\rm ej}v_{\rm ej}^2$, $\tau_{\rm ej}\approx (3-\delta)\kappa M_{\rm ej}/(4\pi R_{\rm ej}^2)$ are respectively the kinetic energy and optical depth of the ejecta with $\kappa$ being the opacity for trapping of photons generated in the wind nebula\footnote{In eq. (\ref{eq:dEraddt}) we assume that the photons carrying the radiative power of the wind nebula and trapped in the ejecta are thermalized. This is because the radiative power from the wind nebula is dominated by inverse-Compton scattering of the thermal UV/optical photons that are trapped in the ejecta, and the scattered photons are in the extreme UV and soft X-ray bands with large absorption cross-sections \citep[see Appendix \ref{sec:radiative power} and also e.g.,][for discussions]{Metzger22}. }. In equation (\ref{eq:dEraddt}), it is assumed that the radiation which is trapped in the ejecta is thermalized as radiation (and not gas). This is a reasonable assumption, as the ejecta's internal energy is dominated by radiation at all times for typical values of temperature and density in our ejecta.

The light curve is set by the luminosity $L_{\rm rad}(t)$ diffusing through the SN ejecta, with the diffusion timescale at time $t$ given as 
\begin{eqnarray}
    t_{\rm diff}(t)&\approx& \zeta\frac{3\kappa M_{\rm ej}}{4\pi c R_{\rm ej}} \nonumber \\
    &\sim& 12\ {\rm day}\left(\frac{\kappa}{0.07\ {\rm cm^2\ g^{-1}}}\right)\left(\frac{M_{\rm ej}}{3~M_\odot}\right)\left(\frac{R_{\rm ej}}{10^{15}\ {\rm cm}}\right)^{-1},
    \label{eq:tdiff_ej}
\end{eqnarray}
where $\zeta\approx 1/3.204$ is an integration factor taking into account the ejecta structure \citep{Arnett80,Arnett82}. For simplicity we fix $M_{\rm ej}$ and the opacity $\kappa$ as constant in time, as the mass of the fast wind merging with the ejecta $M_{\rm w,sh}(\lesssim 0.1~M_\odot)$ is much less than $M_{\rm ej}$. We adopt $\kappa\approx 0.07\ {\rm cm^2\ g^{-1}}$, a reasonable value adopted in stripped-envelope SNe and also inferred for Type I SLSNe \citep[e.g.,][]{Taddia18,Gomez24}.

The heating terms $L_{\rm wind}, L_{\rm Ni}$ are for the disk wind and radioactive decay in the ejecta respectively, with the former in eq. (\ref{eq:Lwind}) and latter modeled as
\begin{eqnarray}
    L_{\rm Ni} &=& [1-\exp(-\tau_{\rm \gamma})] \nonumber \\
    &\times&\left(\frac{M_{\rm Ni}}{M_\odot}\right)\left[L_{\rm Ni,0}\exp\left(-\frac{t}{\tau_{\rm Ni}}\right) + L_{\rm Co,0}\exp\left(-\frac{t}{\tau_{\rm Co}}\right)\right] \nonumber \\
    &+& \left(\frac{M_{\rm Ni}}{M_\odot}\right)L_{\rm Co,1}\left[-\exp\left(-\frac{t}{\tau_{\rm Ni}}\right)+\exp\left(-\frac{t}{\tau_{\rm Co}}\right)\right],
\end{eqnarray}
where $M_{\rm Ni}$ is the mass of $^{56}$Ni in the SN ejecta, $\tau_{\gamma}\approx (3-\delta)\kappa_{\gamma}M_{\rm ej}/(4\pi R_{\rm ej}^2)$ is the gamma-ray trapping depth with $\kappa_\gamma \approx 0.03\ {\rm cm^2\ g^{-1}}$, $\tau_{\rm Ni}=8.8$ days, $\tau_{\rm Co}=111.3$ days, $L_{\rm Ni,0}=6.45\times 10^{43}~{\rm erg\ s^{-1}}$, $L_{\rm Co,0}=1.38\times 10^{43}~{\rm erg\ s^{-1}}$, $L_{\rm Co,1}=4.64\times 10^{41}~{\rm erg\ s^{-1}}$ \citep{Wygoda19}.

\begin{table*}
    \centering
    \begin{tabular}{ccc}
         \hline
         Parameters &  Fiducial values & Range\\
         \hline
         Companion star mass ($M_*$) & $10~M_\odot$ & [3,10,20] $M_\odot$ \\
         Time of star's full disruption ($t_{\rm TDE}$) & 10~{\rm days} & [2,10,30] days\\ 
         Viscosity parameter ($\alpha(H/R)^2$) & $10^{-2}$ & [$3\times 10^{-3}$, $10^{-2}$, $3\times 10^{-2}$] \\
         NS mass, disk inner radius ($M_{\rm NS}, R_{\rm NS}$) & $(1.4~M_\odot, 12~{\rm km}$) & BH model of $M_{\rm BH}=5~M_\odot$ (Section \ref{sec:disk_BH}) \\
         \hline \hline
         SN ejecta parameters & Fiducial ejecta & Other models \\ \hline
         & & HM-1 ($10~M_\odot, 3\times 10^{51}~{\rm erg}, 0.4~M_\odot$) \\
         Mass, kinetic energy, $^{56}$Ni mass ($M_{\rm ej}, E_{\rm exp}$, $M_{\rm Ni}$) & ($3~M_\odot, 10^{51}~{\rm erg}$, $0.15~M_\odot$) & HM-2 ($10~M_\odot, 10^{51}~{\rm erg}, 0.15~M_\odot$)\\
         & & LM ($1~M_\odot, 3\times 10^{50}~{\rm erg}, 0.03~M_\odot$) \\
         Optical/gamma-ray opacities ($\kappa, \kappa_\gamma$) & ($0.07~{\rm cm^2\ g^{-1}}$, $0.03~{\rm cm^2\ g^{-1}}$) & (fixed) \\ \hline
    \end{tabular}
    \caption{Parameters adopted in our model. In addition to varying the parameters governing the disk evolution, we consider the case of a BH remnant (Section \ref{sec:disk_BH}), and four representative cases for the SN ejecta (Section \ref{sec:results}).}
    \label{tab:params}
\end{table*}

The energy injected by the wind is not directly converted to radiation, but is first given to hot ions and electrons in the wind nebula that cool both radiatively via free-free emission and adiabatically. We obtain the efficiency of converting the wind injection to radiation $\epsilon_{\rm rad}$ (eq.  \ref{eq:wind_radiative_efficiency_global}), with the formulations in Appendix \ref{sec:radiative power}. Specifically, we consider cooling of shocked wind gas via free-free emission and inverse Compton scattering off the thermal photons trapped inside the ejecta, using a semi-analytical framework that takes into account the dependence of these processes on the wind velocity.

We set the initial ejecta radius to $R_{\rm ej}(t=0)=R_\odot$ typical for stripped progenitors of Type Ibc SNe, and the initial internal and kinetic energies are assumed to be equally distributed with radiation pressure being dominant, i.e. $E_{\rm rad}=E_{\rm kin}=E_{\rm exp}/2$, $E_{\rm gas}=0$. Here $E_{\rm exp}$ is the explosion energy of the SN ejecta. Our results are insensitive to these initial conditions, as the initial internal energy is quickly converted to kinetic energy by adiabatic expansion on a timescale of $\sim R_{\rm ej}(t=0)/v_{\rm ej}\sim 10^2$ s before any significant heating or radiation plays a role.

The formalism above gives the {\it bolometric} light curve $L_{\rm rad}(t)$, which is the output of radiation trapped in the ejecta and observable as UV-optical-NIR luminosity. In addition to the bolometric light curves, we estimate the $r$-band light curves with bolometric correction \citep{Ofek14}\footnote{Available as a table in SuperNova Explosion Code (\citealt{Morozova15}; \url{https://stellarcollapse.org/index.php/SNEC.html})}, with a solar magnitude of $4.75$. We assume the emission temperature to be the effective temperature $T_{\rm eff}=(L_{\rm rad}/4\pi R_{\rm ej}^2\sigma_{\rm SB})^{1/4}$, where $\sigma_{\rm SB}$ is the Stefan-Boltzmann constant. We further set a temperature floor, that takes into account the effect of the photosphere receding (due to e.g. recombination) as the outer ejecta cools. We adopt a floor value of $6000$ K, reasonable for an optically thick carbon/oxygen-rich ejecta \citep[e.g.,][]{Piro14}. 

We caution that this is a crude estimation and likely less reliable than the bolometric luminosity, especially when the ejecta's optical depth $\tau_{\rm ej}$ falls comparable to unity. When $\tau_{\rm ej}\lesssim$ few, we expect gas near the photosphere departing from local thermal equilibrium, and the color temperature of the emission governed by thermalization is likely higher than $T_{\rm eff}$. When $\tau_{\rm ej}<1$ (nebular phase), our definition of $T_{\rm eff}$ breaks down as the spectra would be highly non-thermal. Obtaining accurate multi-band light curves requires radiative transfer simulations with more detailed opacity treatment, and beyond the scope of this work\footnote{In the SLSN literature, an empirical modified blackbody spectra \citep{Nicholl17} is often adopted, with a linear flux suppression in UV ($F_{\lambda<\lambda_{\rm cut}}=(\lambda/\lambda_{\rm cut})F_\lambda$, $\lambda_{\rm cut}\approx 3000\rm \AA$) and corresponding flux increase at longer wavelengths. Adopting this instead of a pure blackbody brightens the peak $r$-band magnitude by typically $\lesssim 0.3$ mag.}.

\subsection{The case of a black hole remnant}
\label{sec:disk_BH}
We can also consider an analogous case where the remnant is a BH instead of a NS. The main effect of a heavier BH remnant instead of a NS is a higher disk mass and accretion power. The initial disk parameters (eq. \ref{eq:Mdisk_0}, \ref{eq:Rdisk_0}) are updated as
\begin{eqnarray}
    M_{\rm disk,0} &=& M_{\rm BH} \label{eq:Mdisk_0_BH}\\
    r_{\rm disk,0} &=& \frac{2M_{\rm BH}}{M_*}R_* \nonumber\\
    &\sim& 3\times 10^{11}{\rm cm}\left(\frac{M_{\rm BH}}{5~M_\odot}\right)\left(\frac{M_*}{10~M_\odot}\right)^{\!\!-0.4}, \label{eq:Rdisk_0_BH}   
\end{eqnarray}
and the viscous time becomes longer as
\begin{eqnarray}
    t_{\rm visc,0}\sim 5\ {\rm day}\left(\frac{M_{\rm BH}}{5~M_\odot}\right)\left(\frac{M_*}{10~M_\odot}\right)^{\!\!-0.6} \left(\frac{\alpha}{0.1}\right)^{\!\!-1} \left(\frac{H/R}{0.3}\right)^{\!\!-2}.
    \label{eq:t_visc_BH}
\end{eqnarray}
For the case of a BH accretion disk, the inner edge of the disk can be set by the radius of the innermost stable circular orbit, $R_{\rm ISCO}=6GM_{\rm BH}/c^2\approx 44~{\rm km}(M_{\rm BH}/5~M_\odot)$ for a non-spinning BH. The energy budget of the disk wind over the initial viscous time is then
\begin{eqnarray}
    E_{\rm kin,0} &\sim& \frac{1}{2}\frac{GM_{\rm BH}}{R_{\rm ISCO}} \eta_{\rm w}M_{\rm disk,0}\left(\frac{R_{\rm ISCO}}{r_{\rm disk,0}}\right)^{\!\!0.5} \nonumber \\
    &\sim& 1.5\times 10^{51}\ {\rm erg} \left(\frac{M_{\rm BH}}{5~M_\odot}\right) \left(\frac{M_*}{10~M_\odot}\right)^{\!\!0.2},
\end{eqnarray}
potentially capable of powering extreme transients with radiated energy reaching $10^{51}$ erg.

We note though that there are still large uncertainties in the theory of core-collapse SNe forming BHs \citep[e.g.,][]{Sukhbold16,Burrows24} and their natal kicks \citep[e.g.,][]{Repetto12,Mandel16,Atri19,Koshimoto24,Nagarajan24}. Recent simulations \citep{Burrows24} and studies of Galactic BH binaries \citep{Nagarajan24} suggest two broad classes for BH formation: direct collapses with weak kicks ($\lesssim 10$ km s$^{-1}$) and BH-forming SNe with kicks comparable to NSs ($100$--$1000$ km s$^{-1}$). The latter case is more relevant for our scenario, and in this case the mass of the final BH is expected to be lower \citep[$\lesssim 6~M_\odot$; e.g.,][]{Ertl20}.

\section{Results}
\label{sec:results}

\begin{figure*}
     \centering
     \includegraphics[width=\linewidth]{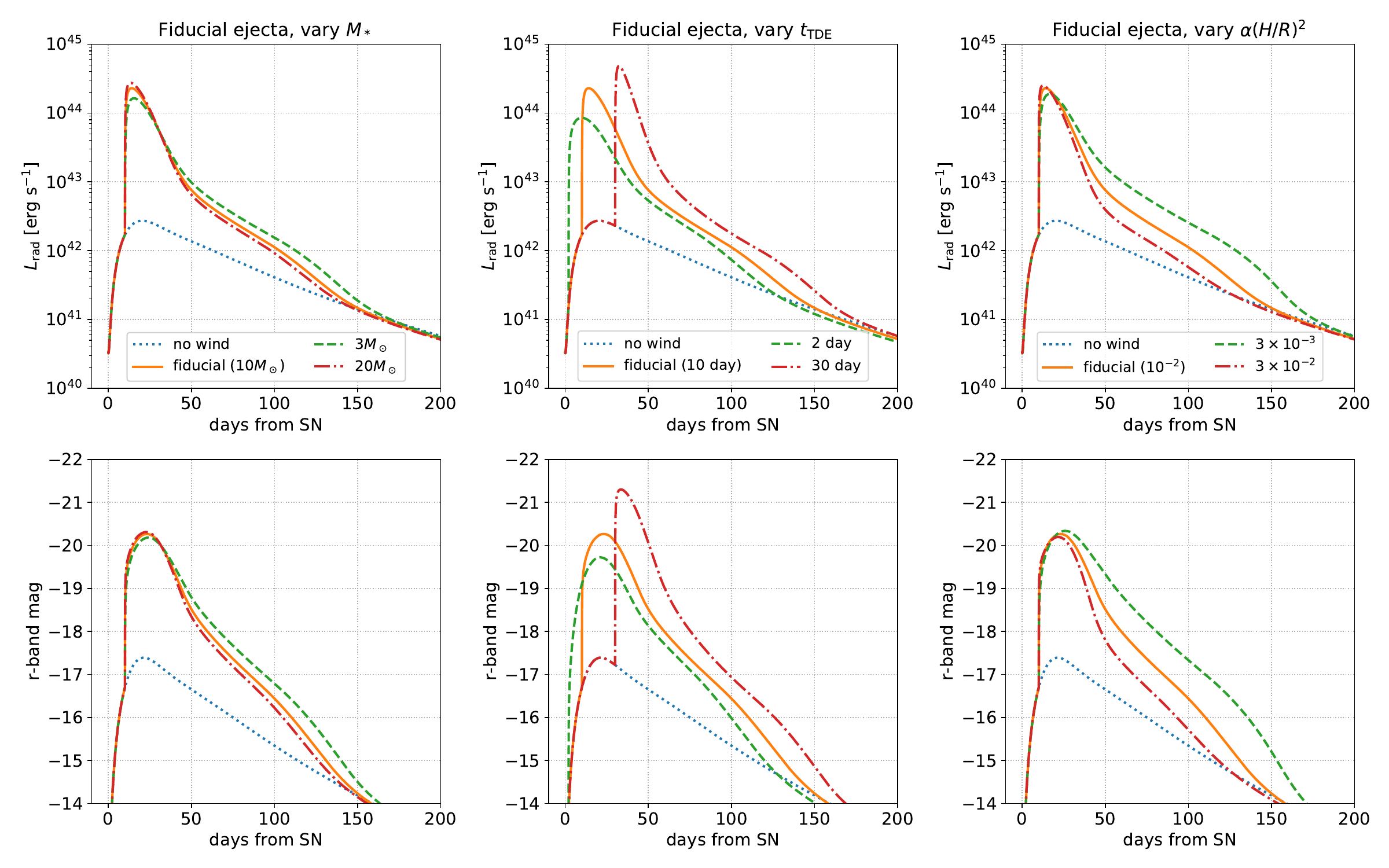}
     \caption{Light curves for the fiducial SN ejecta of $M_{\rm ej}=3~M_\odot$, $E_{\rm ej}=10^{51}$ erg. We consider the fiducial model parameters in Table \ref{tab:params}, and vary each parameter one by one. Top panels are the bolometric light curves, and bottom panels are the $r$-band light curves.}
     \label{fig:lcMej3}
\end{figure*}

Table \ref{tab:params} shows the adopted model parameters. The main free parameters of our model are $t_{\rm TDE}$, $M_*$, and $\alpha(H/R)^2$. The parameter $t_{\rm TDE}$ controls the onset time of the wind injection. The parameters $M_*$ (indirectly through $R_*$) and $\alpha(H/R)^2$ control the viscous timescale, with larger $\alpha(H/R)^2$ and larger $M_*$ leading to faster accretion and energy injection (although the dependence on $M_*$ is weaker, see eq. \ref{eq:t_visc}). For advection-dominated disks we expect $\alpha \sim 0.01$--$0.1$ and $H/R\sim 0.3$--$0.5$, so we consider three values of $\alpha(H/R)^2=(3\times 10^{-3}, 10^{-2}, 3\times 10^{-2}$), with $10^{-2}$ as a fiducial value.

We consider four representative cases of the SN ejecta with ejecta mass, energy and $^{56}$Ni mass of 
\begin{enumerate}
    \item $(M_{\rm ej}, E_{\rm exp}, M_{\rm Ni}) = (3~M_\odot, 10^{51}~{\rm erg}, 0.15~M_\odot)$, that are representative values inferred for Type Ic SNe \citep[e.g.,][]{Lyman16,Taddia18,Rodriguez23}, referred here as the Fiducial model,
    \item $(M_{\rm ej}, E_{\rm exp}, M_{\rm Ni})= (10~M_\odot, 3\times 10^{51}~{\rm erg}, 0.4~M_\odot)$, inferred in a subset ($\sim 10\%$) of SN Ibc with long duration and high ejecta masses \citep{Karamehmetoglu23}, referred here as the HM-1 model,
    \item $(M_{\rm ej}, E_{\rm exp}, M_{\rm Ni}) = (10~M_\odot, 10^{51}~{\rm erg}, 0.15~M_\odot)$, motivated from theoretical modeling of neutrino-driven explosions of high-mass ($\sim 10~M_\odot$) stripped progenitors \citep{Ertl20}, referred here as the HM-2 model.
    \item $(M_{\rm ej}, E_{\rm exp}, M_{\rm Ni}) = (1~M_\odot, 5\times 10^{50}~{\rm erg}, 0.03~M_\odot)$, motivated from  modeling of neutrino-driven explosions of low-mass ($\lesssim$ $3~M_\odot$) stripped progenitors \citep{Ertl20}, referred here as LM model.
\end{enumerate}
The $^{56}$Ni mass for each ejecta model is based on theoretically/observationally motivated values, but the detailed choice of $M_{\rm Ni}$ does not affect the resulting light curves around peak. 

\subsection{Light Curves}
\label{sec:light_curves}
Figure \ref{fig:lcMej3} shows representative light curves for the Fiducial ejecta model of $(M_{\rm ej},E_{\rm exp})=(3~M_\odot, 10^{51}$ erg). The panels are for varying each model parameter $t_{\rm TDE}, M_*$ and $\alpha(H/R)^2$. The bottom panels show the absolute $r$-band light curves, where we find the $r$-band to be in the Rayleigh-Jeans regime for all models around peak.
The timing of energy injection $t_{\rm TDE}$ creates the largest variation in the light curves, due to the photon diffusion time in the ejecta decreasing as it expands. Larger $t_{\rm TDE}$ leads to brighter light curves with shorter rises, as the diffusion time upon the onset of energy injection is shorter. We note that this generally holds in our adopted range of $t_{\rm TDE}$, where $\tau_{\rm ej}\gg 1$ at $t=t_{\rm TDE}$. For very late wind onsets (and/or very low $M_{\rm ej}$) where $\tau_{\rm ej}<1$ already at $t=t_{\rm TDE}$, the peak can instead be dimmer as the efficiency of thermalizing the wind injection (the factor $(1-e^{-\tau_{\rm ej}})\epsilon_{\rm rad}$ in equation \ref{eq:dEraddt}) drops steeply with time for $\tau_{\rm ej}\lesssim 1$.

The other parameters $M_*$, $\alpha(H/R)^2$ have much weaker influence on the light curves, although lower values of $M_*$ or $\alpha(H/R)^2$ leads to slower decay due to the longer viscous timescale (eq. \ref{eq:t_visc}). Increasing $\alpha(H/R)^2$ leads to faster viscous accretion and a brighter bolometric peak, but this is compensated by the smaller $R_{\rm ej}$ (higher $T_{\rm eff}$) and results in similar $r$-band peaks.

\begin{figure*}
    \centering
    \includegraphics[width=\linewidth]{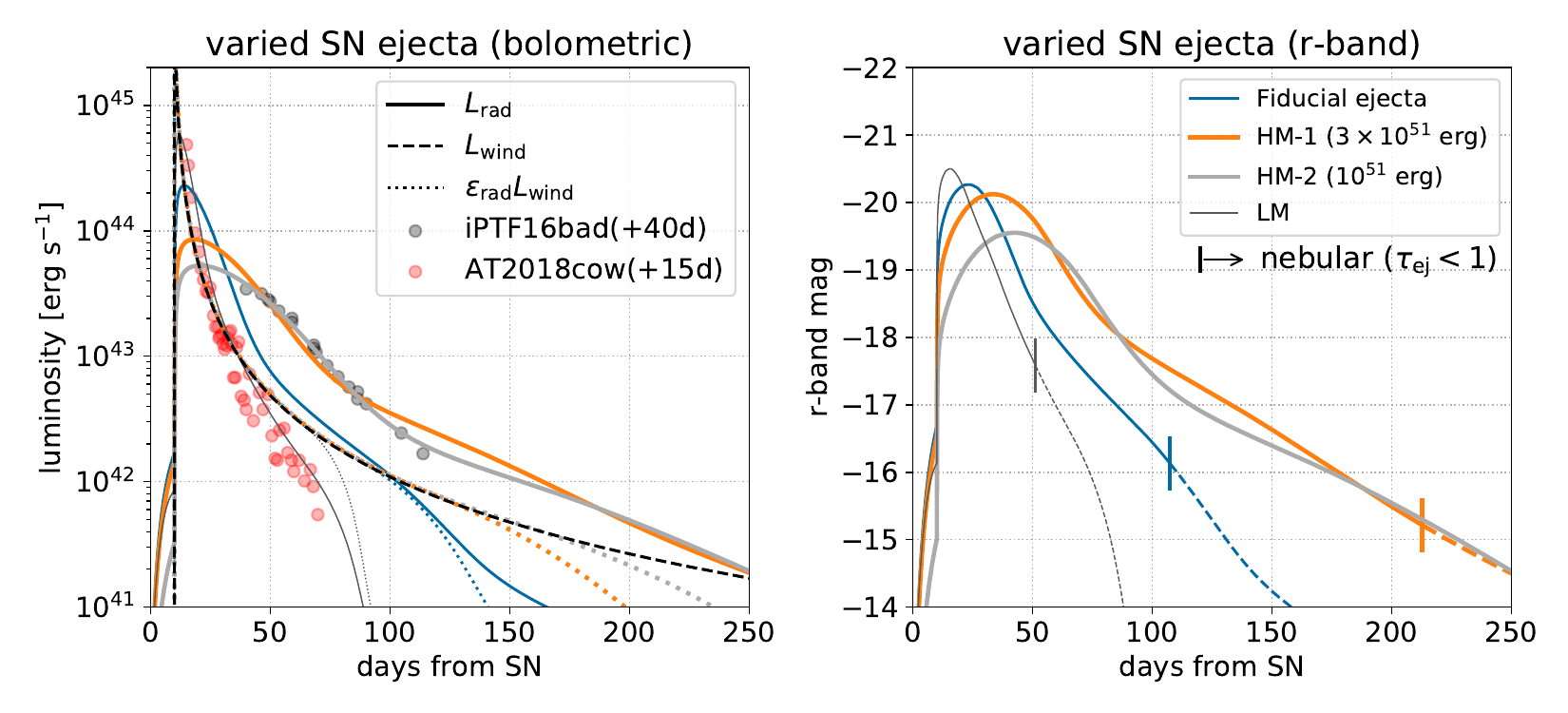}
    \includegraphics[width=\linewidth]{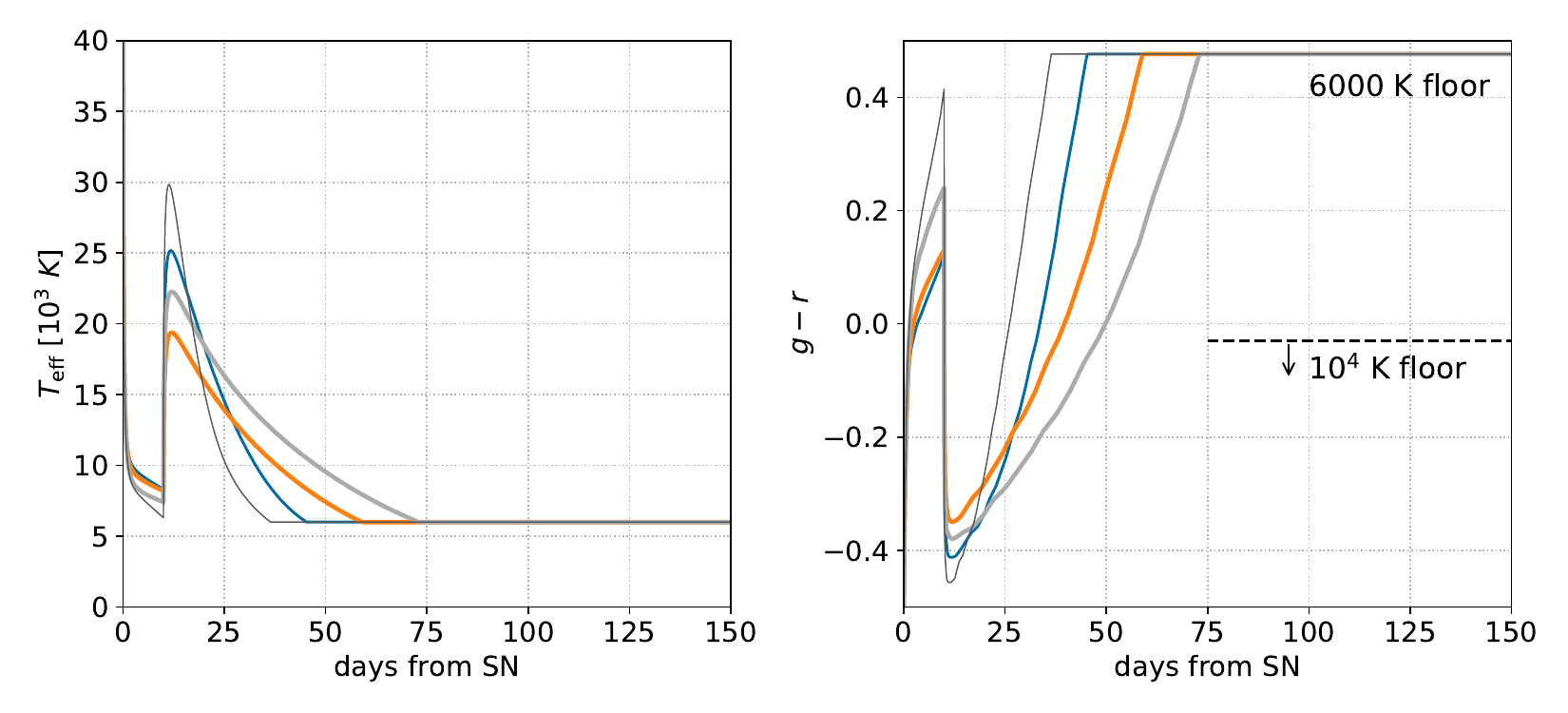}
    \caption{Comparison of the light curves for the four SN ejecta models in Table \ref{tab:params}, shown as solid lines in bolometric (top left panel) and r-band (top right panel). The other parameters in Table \ref{tab:params} are fixed to their fiducial values. Dashed line is the wind injection luminosity $L_{\rm wind}$ (common for all ejecta models), and dotted lines are the wind injection luminosity multiplied by the time-dependent radiation conversion efficiency $\epsilon_{\rm rad}$. We also plot light curves of a SLSN iPTF16bad \citep{Yan17} and an FBOT AT2018cow \citep{Margutti19} that displayed hydrogen line emission in late-time spectra (see Sections \ref{sec:hydrogen}, \ref{sec:fbots}). The bottom panels show evolutions of effective temperature and $g-r$ color for a 6000 K temperature floor. The floor gives a temperature-dependent upper limit on $g-r$, and the dashed line shows that for a floor of $10^4$ K.}
    \label{fig:Mej_comp}
\end{figure*}
Figure \ref{fig:Mej_comp} shows the effects of changing the SN ejecta mass/energy, with other parameters fixed as the fiducial values in Table \ref{tab:params}. The HM-1, HM-2 models evolve on a longer timescale than the Fiducial model over months, while the LM model evolve much faster in days. The main effect of higher $M_{\rm ej}$ and lower $E_{\rm exp}$ is a longer light curve duration due to the longer SN timescale $\propto \sqrt{M_{\rm ej}/v_{\rm ej}} \propto M_{\rm ej}^{3/4}E_{\rm exp}^{-1/4}$. Another effect is to increase the radiation conversion efficiency, as the wind nebula can be confined to smaller radii, where the density of the shocked wind would be higher and the gas cooling would be more efficient. The combined effects create a slower evolution in the light curve.

In Figure \ref{fig:Mej_comp}, we overplot the bolometric light curves of an SLSN iPTF16bad and an FBOT AT2018cow, which are found to display hydrogen lines at late times (see also Sections \ref{sec:hydrogen}, \ref{sec:fbots}). We find that the slow evolution of SLSNe are better reproduced by the high ejecta-mass models (HM-1 and HM-2) with $M_{\rm ej}=10~M_\odot$, while the fast evolution of FBOTs require a much lower ejecta mass of $M_{\rm ej}\lesssim 1~M_\odot$, likely from low-mass helium star progenitors.

Finally in the bottom panels of Figure \ref{fig:Mej_comp} we show the evolutions of the effective temperature $T_{\rm eff}$ and $g-r$ color. We see a quick rise of $T_{\rm eff}$ soon after wind injection, and a drop over weeks to months until it reaches the floor temperature of $6000$ K. Correspondingly $g-r$ quickly drops upon wind injection, and then rises until it plateaus. We note that the value at the plateau is set by the floor temperature, likely dependent on the composition of the ejecta. For a higher floor of $\sim 10^4$ K more appropriate for helium-rich ejecta (that may be expected in LM models; Sec \ref{sec:fbots}), we find $g-r$ of $-0.1$ to $0$, close to the late-time color of AT2018cow \citep{Perley19}.

\subsection{Rise Time and Peak r-band Magnitude}

\begin{figure}
    \centering
    \includegraphics[width=\linewidth]{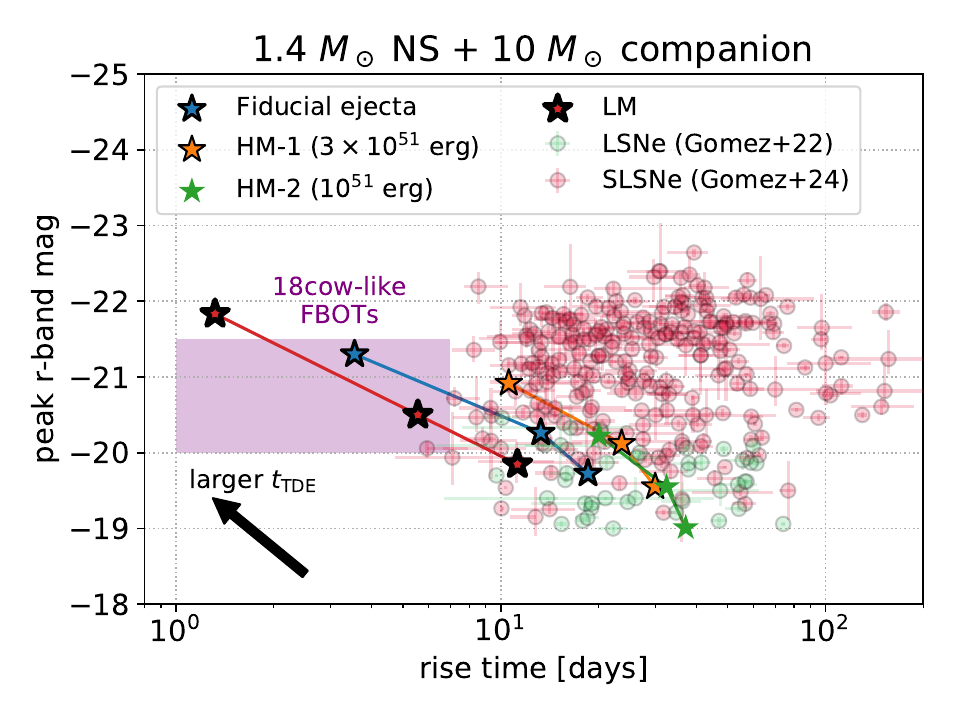}
    \caption{The absolute $r$-band magnitude at peak versus rise time to $r$-band peak. Stars show our NS model with colors separated by the SN ejecta models, with varied $t_{\rm TDE}=[2,10,30]$ days and fixed $M_{\rm *}=10~M_\odot$, $\alpha(H/R)^2=10^{-2}$. The data points show the measurements for samples of luminous and superluminous SNe \citep{Gomez22,Gomez24} brighter than typical SN Ibc. The shaded region shows the rough parameter space of AT2018cow-like FBOTs.
    }
    \label{fig:rise_peak}
\end{figure}

Figure \ref{fig:rise_peak} shows the peak $r$-band magnitude $M_{r, {\rm peak}}$ and the rise time, defined here as the time at $r$-band peak from the onset of wind injection $t=t_{\rm TDE}$. For each SN ejecta model, $t_{\rm TDE}$ are varied while other parameters are fixed as $M_*=10~M_\odot,~\alpha(H/R)^2=10^{-2}$. Within our range of $t_{\rm TDE}$ when $\tau_{\rm ej}\gg 1$ is still satisfied, increasing $t_{\rm TDE}$ makes the peak brighter and decreases the rise time, due to the shorter photon diffusion time at the onset of energy injection. 

These quantities can be compared with the overplotted samples of Type I SLSNe \citep{Gomez24}, and the samples of luminous Type Ibc SNe \citep{Gomez22} defined as a transitional population between normal SN Ibc and Type I SLSNe. For the Fiducial ejecta model of $M_{\rm ej}=3~M_\odot$ with $t_{\rm TDE}\lesssim 10$ days, the peak magnitudes are around -20 mag with rise times of weeks. These most closely overlap with the samples of luminous Type Ibc SNe. 

The Fiducial ejecta model with long $t_{\rm TDE}\gtrsim 10$ days and the LM ejecta model reach a brighter peak of $\approx -21$ mag, but these have timescales of only $\lesssim 10$ days that are shorter than the bulk of the (super-)luminous SN sample. The timescales are more consistent with rapid transients found in high-cadence surveys, in particular the luminous FBOTs with evolution similar to AT2018cow \citep[e.g.,][]{Ho23}.

When compared at the same rise-time, the HM models of $M_{\rm ej} = 10~M_\odot$ can produce peaks brighter than the $3~M_\odot$ model, and closer to the bulk of the SLSN sample. We conclude that massive progenitors ($M_{\rm ej}\gtrsim 10~M_\odot$) would be favorable to produce optical peaks close to $M_{r, \rm peak}\approx -21$ mag and month-long rises typical of SLSNe. This may be consistent with them restricted to low-metallicity environments \citep{Neill11,Perley16} and mostly being helium-poor \citep[][their Table 4]{Woosley19}. However, our fiducial model struggles to explain the brighter and longer events with total radiation energy approaching $10^{51}$ erg, primarily due to the energetics of the wind injection given in eq. (\ref{eq:E_wind}).

\begin{figure*}
    \centering
    \includegraphics[width=\linewidth]{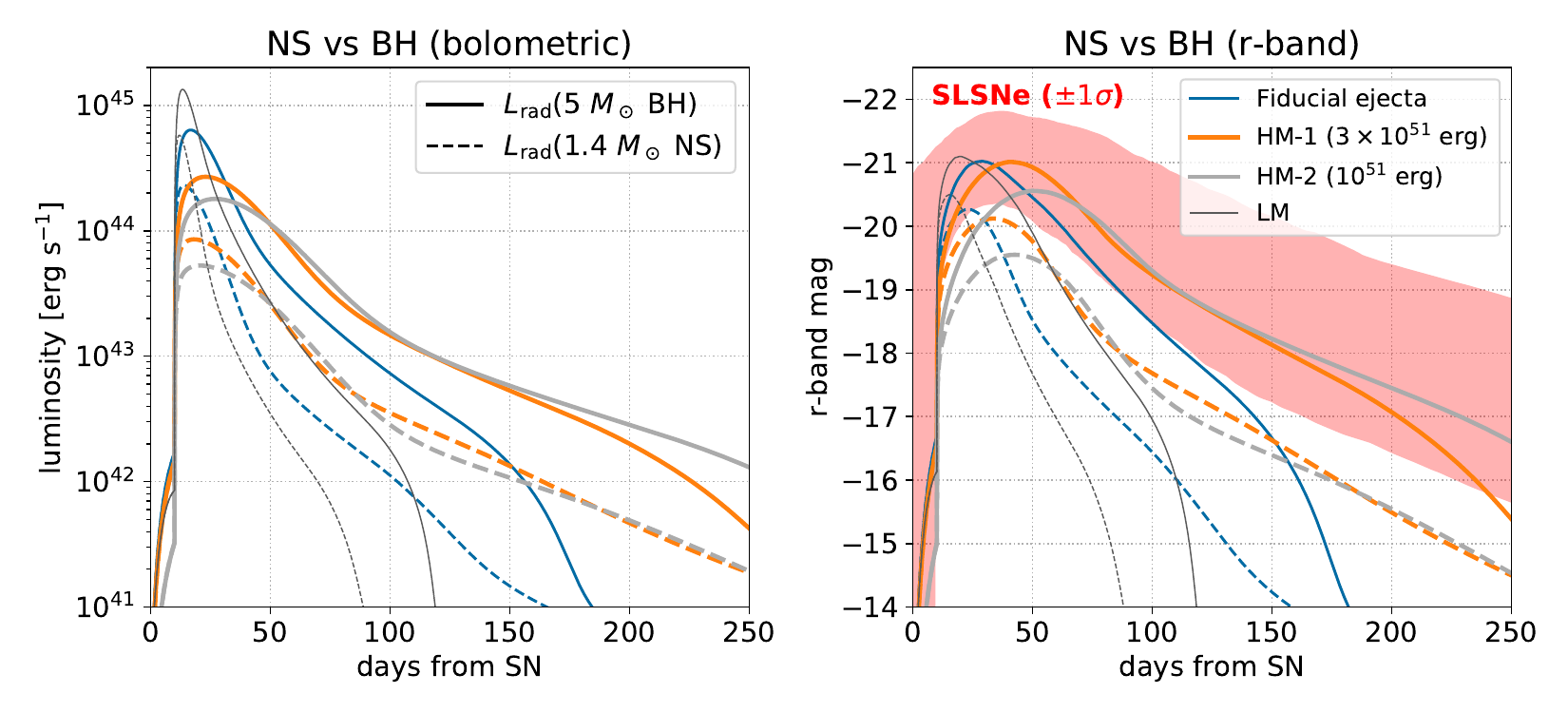}
    \includegraphics[width=\linewidth]{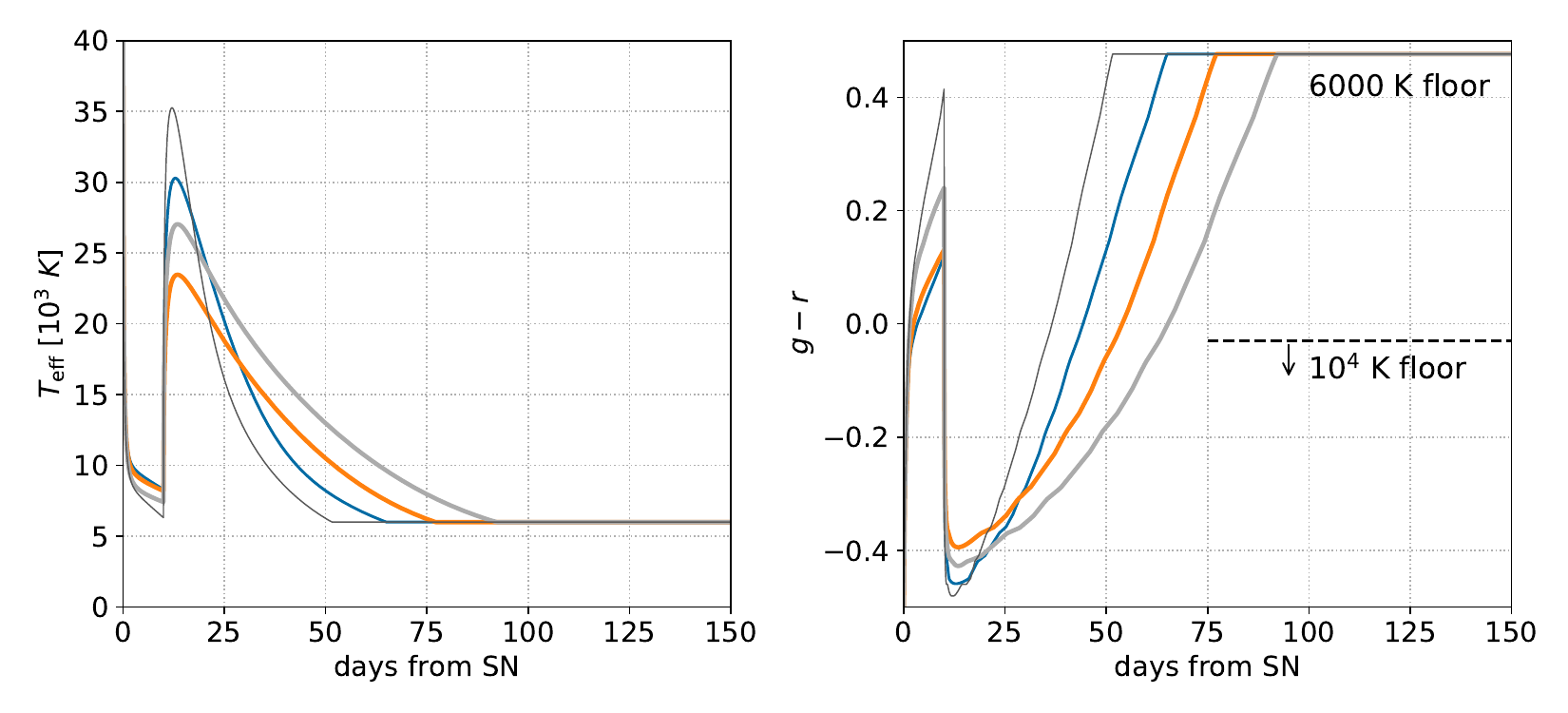}
    \caption{Similar to Figure \ref{fig:Mej_comp}, but a comparison for the case of a BH remnant with mass $M_{\rm BH}=5~M_\odot$ (solid lines) and a NS remnant (dashed lines, same as solid lines in Figure \ref{fig:Mej_comp}). The shaded region in the top right panel shows the $1\sigma$ range for the $r$-band light curves of the SLSN samples in \cite{Gomez24}.}
    \label{fig:BH_remnant}
\end{figure*}

\begin{figure}
    \centering
    \includegraphics[width=\linewidth]{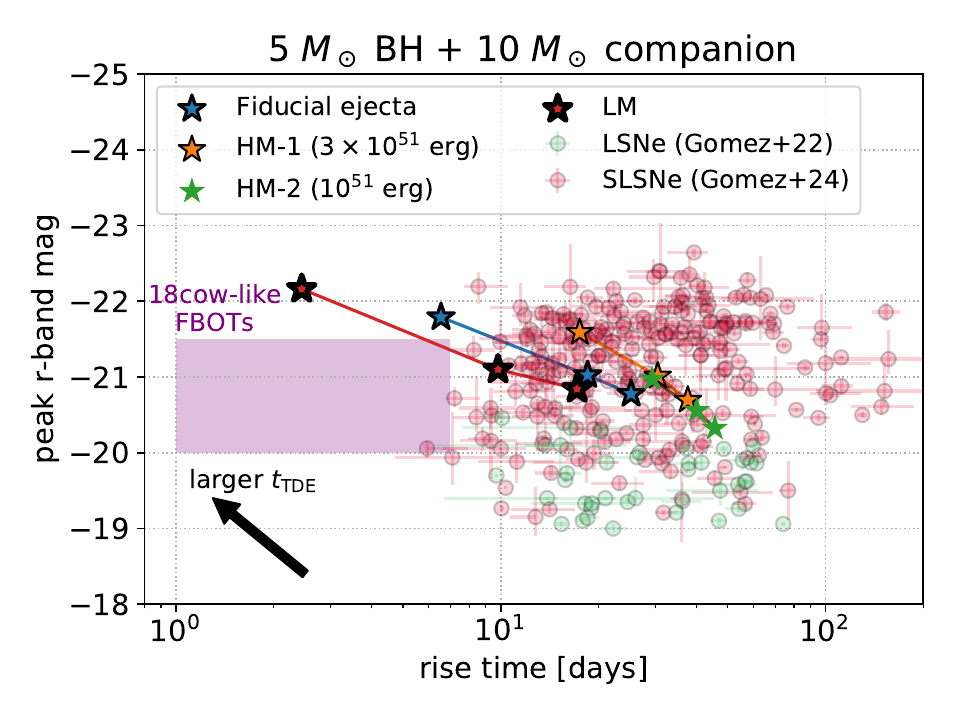}
    \caption{Same as Figure \ref{fig:rise_peak}, but for the case of a $5~M_\odot$ BH remnant disrupting the $10~M_\odot$ main-sequence star. The peaks are generally $\approx 1$ mag brighter than for a NS remnant.}
    \label{fig:rise_peak_BH}
\end{figure}

Explaining brighter SLSNe of $M_{\rm r, peak}\lesssim -21$ mag by our model requires enhancements in the energy budget of the disk wind. Three possible extensions/modifications to our model may realize this: (1) a BH remnant instead of a NS, (2) multiple encounters that can prolong and enhance the accretion onto the NS; and (3) a smaller accretion rate power-law index $p$ than what has been adopted in our model \citep[$p=0.5$, motivated by][]{Cho24_ADAF_accretion_profile, guo24_ADAF_accretion_profile}. The uncertainty in $p$ affects both the accretion power and the disk evolution, and more theoretical works along the direction of \citet{stone99_ADAF_wind, Yuan12, Cho24_ADAF_accretion_profile, guo24_ADAF_accretion_profile} are needed to better understand the behaviors of radiatively inefficient disks on timescales much longer than the initial viscous time. While we leave the possibility of the latter two to future work (see Section \ref{sec:multiple_encounter} for qualitative discussions on multiple encounters), we calculate the effect of the first case of a BH remnant following the prescriptions in Section \ref{sec:disk_BH}.

Figure \ref{fig:BH_remnant} shows light curves comparing the cases of a $5~M_\odot$ BH (solid lines) and a NS remnant (dashed lines), and Figure \ref{fig:rise_peak_BH} shows the rise time vs. peak magnitude relation for the $5~M_\odot$ BH cases. A heavier remnant increases both the disk mass and viscous time, which makes the light curve brighter at all times by $\approx 1$ mag. Brighter SLSNe of $M_{r, {\rm peak}}=-21$ to $-22$ mag (shaded region in Figure \ref{fig:BH_remnant}; \citealt{Gomez24}) are best reproduced in our model by the HM ejecta (and possibly Fiducial ejecta) leaving a kicked BH remnant.

Energy injection by spindown of magnetars, with initial periods of milliseconds, has been popularly invoked as explanation for Type I SLSNe \citep[e.g.,][]{Kasen10,Woosley10}. Recent work on modeling tidal spin-up of their helium star progenitors finds that only massive progenitors of mass $\gtrsim 10~M_\odot$ can produce NSs with spin periods of $\lesssim 5 $ ms required to reproduce the light curves of SLSNe \citep[][their Figure 5]{Fuller22}. If this is correct, the magnetar model is thus energetically successful at explaining SLSNe from massive progenitors with $M_{\rm ej}\gtrsim 10~M_\odot$ that predict long bright light curves, but has a potential difficulty on explaining faster events with lower ejecta masses, say $M_{\rm ej}\lesssim 5~M_\odot$. While the magnetar model remains a more plausible explanation for the brightest and longest events in the SLSN sample, our model may explain a potentially dominant fraction of the relatively dimmer ($\gtrsim -21$ mag) and faster (rise times of $\lesssim$ month) SLSNe, with low ejecta masses inferred in the magnetar model.

Finally, we note the similarity of our bolometric light curves with a population of fast-cooling bright transients with peak magnitudes of $\approx -22$ mag and optical rise times of $\lesssim 10$ days \citep{Nicholl23}. Association of these events with passive galaxies at large nuclear offsets raised a scenario of TDEs involving a BH and a low-mass star in dense clusters \citep[e.g.,][]{Kremer23}, but the strong adiabatic loss in the disk wind results in luminosities much dimmer in the optical. Repeated encounters in these systems may produce brighter emission by wind-ejecta collisions like our model, although we leave detailed light curve modeling to future work.

\section{Discussion}
\label{sec:discussion}
\subsection{Event Rates}
\label{sec:rates}
Motivated by the diverse light curve properties of the tidal disruptions and the classes of transients they may reproduce, we consider the expected rates of these disruptions via Monte-Carlo analysis. We solve the orbit of the binary after a SN explosion of the stripped progenitor, under an isotropic natal kick of the NS/BH remnant.

In Section \ref{sec:model} we have adopted a criterion of the pericenter distance as $r_{\rm p}<R_*$ for a disruption of the star. Here we generalize $r_{\rm p}$ to the distance at closest approach $r_{\rm cl}$ of the post-SN binary. The value of $r_{\rm cl}$ is equivalent to $r_{\rm p}$ for bound post-SN orbits, where the binary is guaranteed to return to the closest approach. Even for unbound orbits under the point-mass approximation, the NS may be kicked towards the companion and have an orbit intersecting the companion. In either case, we can obtain $r_{\rm cl}$ by solving for the post-SN orbit as follows.

We assume that the pre-SN binary is in a circular orbit with semimajor axis $a_{\rm bin}$, composed of a stripped progenitor of mass $M_{\rm prog}$ and a companion of mass $M_*$. After a SN of ejecta mass $M_{\rm ej}$, the helium star is assumed to leave a compact object of mass $M_\bullet=M_{\rm prog}-M_{\rm ej}$, either a NS of mass $M_{\rm \bullet}=1.4~M_\odot$ or a BH of mass $M_\bullet=5~M_\odot$, that receives a natal kick of magnitude $v_{\rm kick}$ at a random orientation. For a NS remnant we assume that the kick speed follows a Maxwell-Boltzmann distribution
\begin{eqnarray}
    p\left(v_{\rm kick}|\sigma_{\rm kick}\right) = \sqrt{\frac{2}{\pi}}\frac{v_{\rm kick}^2}{\sigma_{\rm kick}^3}\exp\left(-\frac{v_{\rm kick}^2}{2\sigma_{\rm kick}^2}\right),
\end{eqnarray} 
with $\sigma_{\rm kick}=265\ {\rm km\ s^{-1}}$ \citep{Hobbs05}. The kicks for BH remnants are less certain, and we assume a log-uniform distribution from 10 km s$^{-1}$ to 2000 km s$^{-1}$, based on the range of values found in recent studies \citep[][see also Section \ref{sec:disk_BH}]{Burrows24,Nagarajan24}.

We place the binary as in Figure 1 of \cite{Brandt95}, with the orientation of the kick set by the random variables $\phi$ ($0\leq\phi<2\pi$) and $\theta$ ($-\pi/2\leq \theta\leq \pi/2$). We can obtain the closest approach of the NS/BH and the main sequence star, by solving the evolution of the displacement vector between the two objects
\begin{eqnarray}
    \frac{d^2\vec{r}_{12}}{dt^2} = -\frac{G(M_{\rm \bullet}+M_*)}{|\vec{r}_{12}|^3}\vec{r}_{12},
    \label{eq:d2rdt2}
\end{eqnarray}
with initial conditions as $x_{12}=z_{12}=0, y_{12}=a_{\rm bin}$ and
\begin{eqnarray}
    v_{x,12} &=& \sqrt{G(M_{\rm prog}+M_*)/a_{\rm bin}} + v_{\rm kick}\cos\theta\cos\phi\\
    v_{y,12} &=& v_{\rm kick}\cos\theta\sin\phi\\
    v_{z,12} &=& v_{\rm kick}\sin\theta.
\end{eqnarray}
For bound orbits $r_{\rm cl}$ corresponds to the pericenter radius $r_{\rm p}$, and an analytical solution for the final orbital parameters exists \citep{Brandt95}. Using two dimensionless quantities
\begin{eqnarray}
    \tilde{m}=\frac{M_{\rm prog}+M_*}{M_{\bullet}+M_*},~ \tilde{v} = \frac{v_{\rm kick}}{v_{\rm orb}},
\end{eqnarray}
$r_{\rm cl}=r_{\rm p}=a_{\rm post}(1-e_{\rm post})$, where the post-SN semimajor axis and eccentricity are
\begin{eqnarray}
    a_{\rm post} &=& \frac{1}{2-\tilde{m}[1+2\tilde{v}\cos\phi \cos\theta + \tilde{v}^2]}a_{\rm bin}\\
    e_{\rm post}^2 &=& 1-\tilde{m}\left\{2-\tilde{m}[1+2\tilde{v}\cos\phi \cos\theta + \tilde{v}^2]\right\} \nonumber \\
    &&\times \left[(1+\tilde{v}\cos\phi\cos\theta)^2+(\tilde{v}\sin\theta)^2\right].
\end{eqnarray}
For bound orbits (i.e. $a_{\rm post}>0$), we use this analytical solution to estimate $r_{\rm cl}$, and for unbound orbits we numerically solve $\vec{r}_{12}$ by eq. (\ref{eq:d2rdt2}) for 100 initial orbital periods and find the minimum value of $|\vec{r}_{12}|$. 

\begin{figure*}
    \centering
    \includegraphics[width=\linewidth]{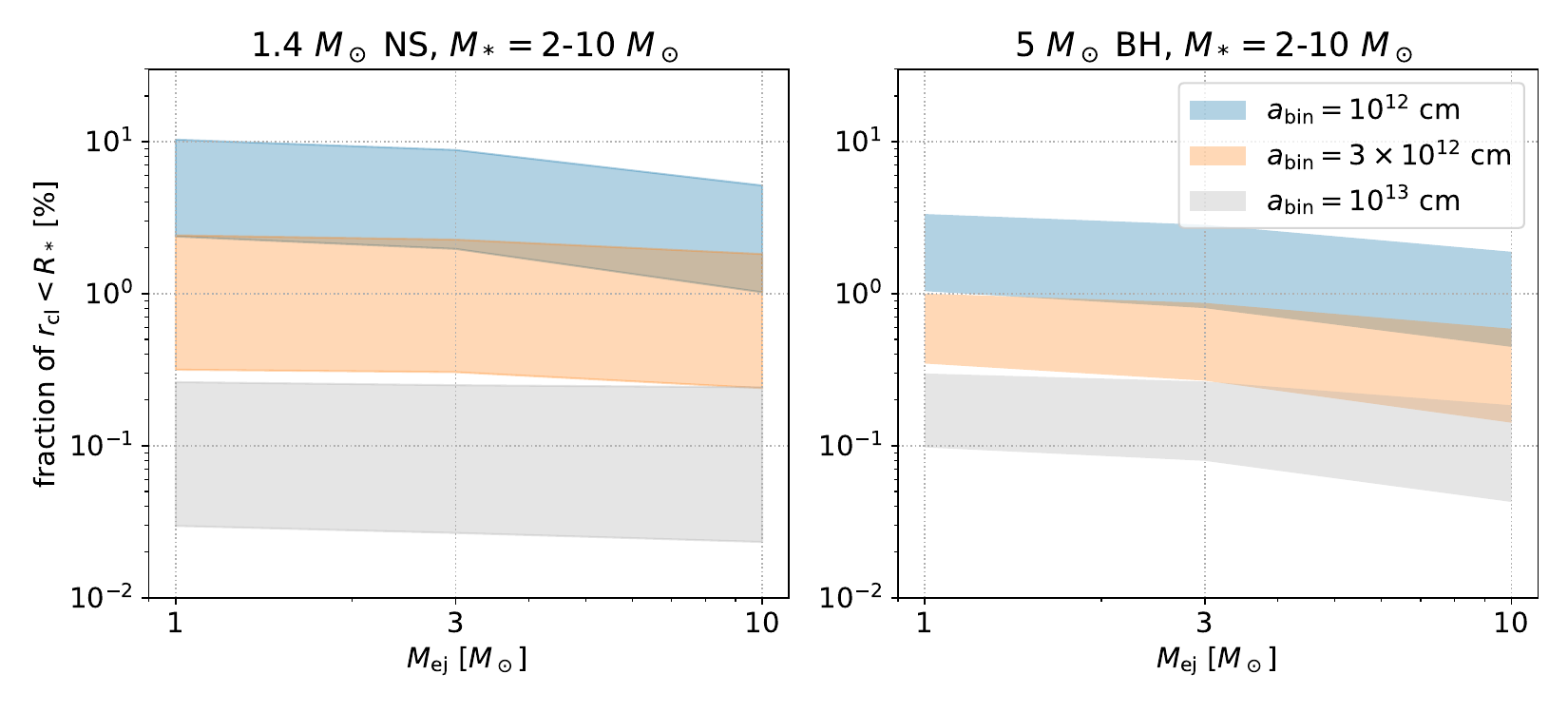}
    \caption{Fraction of SNe where the companion star is tidally disrupted by the newborn NS/BH under our criterion. The two panels show cases for a $1.4~M_\odot$ NS and $5~M_\odot$ BH remnant, with different prescriptions for the natal kick distribution (see main text). Bands indicate the range for the companion mass of $M_*=2$--$10~M_\odot$, with colors varied by the pre-SN separation $a_{\rm bin}$.}
    \label{fig:rp_MC}
\end{figure*}

Figure \ref{fig:rp_MC} shows the fraction of SNe where our adopted criterion of $r_{\rm cl} < R_*$ is satisfied, for 3 cases of $M_{\rm ej}=[1,3,10]~M_\odot$ and $a_{\rm bin}=[1,3,10]\times 10^{12}$ cm. Each band corresponds to the range of $M_*=2$--$10~M_\odot$, with the disruption probability higher for larger $M_*$. We generally predict a probability of $0.03$--$10$\%, with strong dependence on $a_{\rm bin}$, $M_*$ and weaker dependence on $M_{\rm ej}$.

A lower limit of $a_{\rm bin}$ for the main-sequence companion to not be overfilling its Roche lobe \citep{Eggleton83} is $a_{\rm bin}\gtrsim$ ($0.3$--$1$)$\times 10^{12}$ cm, for companion masses $M_*=2$--$20~M_\odot$ and $M_{\rm prog}=5~M_\odot$ (with weak dependence on $M_{\rm prog}$). Tight binary separations of $a_{\rm bin}\lesssim 10^{12}$ cm (corresponding to orbital periods of $\lesssim$ 1--2 days), if sustained from earlier in their evolution, would likely have resulted in stellar mergers and not have produced a detached binary at the time of SN \citep[e.g.,][]{deMink13,Kinugawa24}. Therefore we expect the pre-SN binary to likely have separations of $a_{\rm bin}\gtrsim$ (a few)$\times 10^{12}$ cm, as also found in detailed binary population modeling \citep{Moriya15}. For ranges of $a_{\rm bin}=(3$--$10)\times 10^{12}$ cm and $M_*=5$--$10~M_\odot$ typically found from binary population synthesis models \citep{Moriya15,Zapartas17}, we find the TDE fraction to be $\sim 0.1$--$2\%$.

The rates of these disruptions can be roughly compared with the rates of transients estimated from optical surveys. The rate of luminous Ibc brighter than -20 mag are estimated to be $\sim 1\%$ of SN Ibc \citep{DeCia18}, with the brightest SLSNe being $\sim 0.1\%$ \citep[][and references therein]{Frohmaier21}. AT2018cow-like FBOTs are also very rare, with upper limits of $<(0.1$--$0.4)~\%$ of core-collapse SNe \citep{Coppejans20,Ho23}, i.e. $\lesssim 0.3$--$1~\%$ of SN Ibc. While the ranges of our estimate are too large for detailed comparisons, they are roughly capable of explaining the event rates of these transients. Our estimates can be improved in future works with detailed predictions of $(a_{\rm bin}$, $M_*)$ and SN physics from e.g. population synthesis modeling.

\subsection{Luminous SNe Ibc with Late-time Hydrogen Lines}
\label{sec:hydrogen}
Nebular spectra of Type I SLSNe at $\gtrsim 100$ days from peak were reported in several studies \citep{Yan15,Yan17,Nicholl19}. \cite{Yan17} reports a non-negligible fraction ($\sim 10$--$30\%$) of SLSNe having late-time H$\alpha$ emission with luminosities of $L_{\rm H \alpha}\sim$(0.5--3)$\times 10^{41}$ erg s$^{-1}$ and widths of several $1000$ km s$^{-1}$. Similar late time H$\alpha$ emission was also tentatively suggested in a luminous Type Ic SN 2012aa \citep{Roy16}. 

The late-time hydrogen features with such velocities would be naturally expected from our model, where the hydrogen-rich material from the disrupted companion is responsible for the H$\alpha$ emission once the hydrogen-poor SN ejecta becomes optically thin. The bulk of the hydrogen rich material is confined within the wind nebula, so the expansion velocity of the wind nebula (see Figure \ref{fig:nebula_prop} and end of Section \ref{sec:nebula}) sets a rough estimate for the velocity dispersion of the hydrogen line of several $10^3$ km s$^{-1}$, consistent with the observed line width.

There are two questions in the context of our scenario that we believe are less trivial. First, what mechanism sets the observed H$\alpha$ luminosity? Second, would it be consistent with the H$\alpha$ line not observed at earlier times around light curve peak? We discuss these questions below.

For the first question, copious H$\alpha$ photons can be generated by recombination of hydrogen in the slower disk wind photo-ionized by the hard radiation from the wind nebula, with luminosity $L_{\rm rad}\sim 10^{42}$--$10^{43}$ erg s$^{-1}$ at late times. From Figure \ref{fig:disk_evolve}, at late phases after several viscous times (eq. \ref{eq:t_visc}) a majority of the bound material from the disrupted star has been lost into the un-shocked \textit{slow} disk wind, with mass of $M_{\rm wind,sl}\sim M_{\rm disk,0}$ confined within the nebula radius
\begin{eqnarray}
    R_{\rm neb} &\approx& t\frac{dR_{\rm neb}}{dt}\nonumber \\
    &\approx& 3.5\times 10^{15}\ {\rm cm}\left(\frac{dR_{\rm neb}/dt}{4000~{\rm km\ s^{-1}}}\right)\left(\frac{t}{100\ {\rm day}}\right).
\end{eqnarray}
Assuming ionization balance, the mass of the wind that is photo-ionized is estimated as
\begin{eqnarray}
    M_{\rm ion}\approx {\rm min}\left(M_{\rm wind, sl}, \frac{\dot{N}_{\rm ion} m_p}{n_e \alpha_B}\right),
\end{eqnarray}
where $m_p$ is the proton mass, $n_e\gtrsim 3M_{\rm ion}/(4\pi R_{\rm neb}^3 m_p)$ is the free electron number density (given that these electrons are located within $R_{\rm neb}$), $\alpha_{\rm B}\approx 2.6\times 10^{-13}(T/10^4\ {\rm K})^{-0.7}\ {\rm cm^3\ s^{-1}}$ is the case-B recombination coefficient of hydrogen, and $\dot{N}_{\rm ion}$ is the total ionization rate given by
\begin{eqnarray}
    \dot{N}_{\rm ion}&\approx& \xi\frac{L_{\rm rad}}{\epsilon_{\rm ion}} \nonumber \\
    &\sim& 10^{53}\ {\rm s^{-1}}\left(\frac{\xi}{0.5}\right) \left(\frac{L_{\rm rad}}{10^{43}\ {\rm erg\ s^{-1}}}\right)\left(\frac{\epsilon_{\rm ion}}{30\ {\rm eV}}\right)^{-1}.
\end{eqnarray}
Here we assumed a fraction $\xi\sim 0.5$ of the luminosity $L_{\rm rad}$ produced by the wind nebula is directed inwards and used to ionize the inner wind, with characteristic energy cost of $\epsilon_{\rm ion}$ for each hydrogen ionization. While obtaining $\epsilon_{\rm ion}$ requires solving the processes of photo-ionization and recombination for each species in the wind, hereafter we adopt $\epsilon_{\rm ion}\sim 30$ eV, a reasonable estimate when including the ionization potential of neutral hydrogen/helium and the kinetic energies of the free ions and electrons.

If the entire unshocked slow wind is ionized, i.e., $M_{\rm ion} = M_{\rm wind,sl}$ (in the so-called density-bounded regime; \citealt{osterbrock2006}), then the H$\alpha$ luminosity is given by the total recombination rate of the entire unshocked wind and hence does not track the ionizing luminosity.

On the other hand, if only part of the unshocked wind is ionized ($M_{\rm ion} < M_{\rm wind,sl}$; in the so-called ionization-bounded regime), the H$\alpha$ luminosity is simply set by the ionization rate $\dot{N}_{\rm ion}$ under ionization balance, and the H$\alpha$ luminosity is expected to track the ionizing luminosity. From the aforementioned lower limit on $n_e$, we obtain a constraint on the mass of ionized hydrogen
\begin{eqnarray}
    M_{\rm ion}&<&m_p\sqrt{\frac{4\pi R_{\rm neb}^3 (\xi L_{\rm rad}/\epsilon_{\rm ion})}{3\alpha_B}} \nonumber\\
    &\lesssim& 0.3\ M_\odot \left(\frac{\xi}{0.5}\right)^{0.5}\left(\frac{L_{\rm rad}}{10^{43}\ {\rm erg\ s^{-1}}}\right)^{0.5} \nonumber \\
    &&\times \left(\frac{\epsilon_{\rm ion}}{30\ {\rm eV}}\right)^{-0.5}\left(\frac{R_{\rm neb}}{4\times 10^{15}\ {\rm cm}}\right)^{1.5}\left(\frac{T}{10^4\ {\rm K}}\right)^{0.35},
    \label{eq:Mion_max}
\end{eqnarray}
which indicates our system to be most likely in the ionization-bounded regime. Nevertheless, the hydrogen column density in the ionized region $N_{\rm H} \sim 10^{24}\mr{\,cm^{-2}} (M_{\rm ion}/0.3M_\odot) (R_{\rm neb}/4\times10^{15}\mr{\,cm})^{-2}$ is already so large that the incoming high-energy (X-ray) photons will likely be reprocessed to lower energy photons with energies comparable to $\epsilon_{\rm ion}$ when reaching the inner neutral wind. Assuming that H$\alpha$ photons originates from recombination emission of hydrogen ionized by these photons, its luminosity is
\begin{eqnarray}
    L_{\rm H\alpha} &\sim& \xi\frac{L_{\rm rad}}{\epsilon_{\rm ion}} \frac{\alpha^{\rm H\alpha}_{\rm B}}{\alpha_{\rm B}}\epsilon_{\rm H\alpha} \nonumber \\
    &\sim& 10^{41}\ {\rm erg\ s^{-1}} \left(\frac{\xi}{0.5}\right) \left(\frac{L_{\rm rad}}{10^{43}~{\rm erg\ s^{-1}}}\right)\left(\frac{\epsilon_{\rm ion}}{30\ {\rm eV}}\right)^{-1},
    \label{eq:L_Halpha}
\end{eqnarray}
where $\epsilon_{\rm H\alpha}\approx 1.9~{\rm eV}$ is the energy of the H$\alpha$ photon, and $\alpha^{\rm H\alpha}_{\rm B}/\alpha_{\rm B}\approx 0.3$ \citep{osterbrock2006} is the branching ratio for H$\alpha$ emission in the case-B limit that is weakly dependent on temperature.

We note that the estimation in eq. (\ref{eq:L_Halpha}) does not depend on the exact value of $R_{\rm neb}$ or the electron density $n_e$ in the ionized region, and the H$\alpha$ luminosity tracks the bolometric luminosity $L_{\rm rad}$ in the ionization-bounded regime with a roughly constant ratio of $L_{\rm H\alpha}/L_{\rm rad}\sim 1\%$. For two of the three samples of \cite{Yan17} where bolometric light curve data is available at late times, the detected H$\alpha$ luminosity drops with time with a slope similar to the bolometric light curve, roughly consistent with the prediction from this model. 

For the second question, the H$\alpha$ emission from the unshocked wind is initially smeared due to Compton scattering by the SN ejecta embedding the wind. The scattering optical depth of the ejecta is obtained from integrating eq. (\ref{eq:rho_ej}) as
\begin{eqnarray}
    \tau_{\rm scat}&\approx& \frac{\kappa}{2\pi}\frac{M_{\rm ej}}{R_{\rm ej}^2}\ln\left(\frac{R_{\rm ej}}{R_{\rm neb}}\right) \nonumber \\
    &\sim& 1\left(\frac{\kappa}{0.07\ {\rm cm^2\ g^{-1}}}\right)\left(\frac{M_{\rm ej}}{3~M_\odot}\right)\left(\frac{R_{\rm ej}}{8\times 10^{15}\ {\rm cm}}\right)^{-2},
\end{eqnarray}
where we approximated $\ln(R_{\rm ej}/R_{\rm neb})\approx 1$ (Figure \ref{fig:nebula_prop}). When $\tau_{\rm scat}\gtrsim 1$, Compton scattering creates broad scattering wings that smear the line emission. As the H$\alpha$ luminosity is $\sim 1\%$ of the bolometric continuum luminosity (eq. \ref{eq:L_Halpha}), we expect the emission line to stand out from the continuum only when its width is sufficiently narrow $\lesssim 0.01c$. Due to the ejecta having a bulk velocity larger than $0.01c$, we expect the line emission to be detectable when the scattering optical depth drops to $\tau_{\rm scat}\lesssim 1$. Using $R_{\rm ej}\approx v_{\rm ej}t$ and $v_{\rm ej}=\sqrt{10E_{\rm ej}/3M_{\rm ej}}$, this condition is satisfied from 
\begin{eqnarray}
    t_{\rm H\alpha}&\sim& 130 \ {\rm day} \left(\frac{\kappa}{0.07\ {\rm cm^2\ g^{-1}}}\right)^{\!\!1/2}\nonumber \\
    &&\times \left(\frac{M_{\rm ej}}{3~M_\odot}\right)\left(\frac{E_{\rm ej}}{10^{51}\ {\rm erg}}\right)^{\!\!-1/2},
\end{eqnarray}
i.e. roughly $100$--$400$ days from explosion for the Fiducial and HM ejecta models. This weakly depends on our simplified one-zone treatment of the scattering opacity $\kappa$ that in reality can vary within the ejecta, with the inner region having higher $\kappa$ as it is continuously photo-ionized by the wind nebula.
Our estimate of $t_{\rm H\alpha}$ explains the H$\alpha$ line being detected at late times, but not detected at early times at around the light curve peak.

\subsection{Multi-wavelength Observations of FBOTs}
\label{sec:fbots}

In recent years, high-cadence surveys have found a population of (luminous) FBOTs with peaks of $-21$ mag ($\sim 10^{44}$ erg s$^{-1}$) and evolution timescale of days. Extensive multi-wavelength follow-up of the prototype event AT2018cow have been carried out \citep{Prentice18,Perley19,Margutti19}. The observations of AT2018cow indicate the presence of fast outflow expanding at $\sim 0.1c$ \citep{Margutti19,Ho19}, a central source bright in X-rays at early times and in UV at late times \citep{Margutti19,Sun22,Chen23,Migliori24}, and a low $^{56}$Ni yield of $M_{\rm Ni}\lesssim 0.05$--$0.1~M_\odot$ \citep{Perley19,Margutti19}. 

In our model, such observations are best reproduced by an accreting NS formed from an explosion of a low-mass ($\lesssim 3~M_\odot$) helium star, which likely lost its hydrogen-rich envelope through previous binary interaction. Such stars are expected to explode with a low ejecta mass of $M_{\rm ej}\lesssim 1~M_\odot$, depending on the strength of further stripping via Case BC mass transfer \citep[e.g.,][]{Wu22,Ercolino24}. 

Intriguingly, FBOTs with late-time spectral coverage (AT2018cow, CSS161010) show spectral transitions analogous to that discussed in Section \ref{sec:hydrogen}. The spectra are initially featureless, but emission lines of hydrogen and helium appear at $\approx 20$ days from peak, with line widths of $4000$--$6000$ km s$^{-1}$ in AT 2018cow \citep{Perley19,Margutti19} and 4000--10000 km s$^{-1}$ in CSS161010 \citep{Gutierrez24}. These velocities are consistent with the slower disk wind embedded in the SN ejecta, and the earlier time of spectral transition with respect to SLSNe can be explained by the much lower $M_{\rm ej}$ expected for these systems. It should be noted that, for lower ejecta masses $M_{\rm ej}\lesssim 1M_\odot$ and faster expansion speeds, the radius of the wind nebula $R_{\rm neb}$ expands faster than in the case of SLSNe. For this reason, we expect the ionization of the slower unshocked disk wind to be in the density-bounded regime for FBOTs, and hence the ratio between the H$\alpha$ line luminosity and the bolometric luminosity is likely lower than SLSNe ($L_{\rm H\alpha}/L_{\rm rad}<1\%$) and does not stay constant over time. As the system evolves towards deeper in the density-bounded regime, the H$\alpha$ yield $L_{\rm H\alpha}/L_{\rm rad}$ should drop over time. We leave a detailed modeling of the line evolution to a future work \citep[along the direction of][]{dessart20_Ia_embed_Hrich_gas}.

At a similar timing with the spectral transition, AT 2018cow also showed a transition in X-rays with respect to optical/UV. Initially the X-rays were sub-luminous compared to the optical, but became comparable from around $\approx 20$ days. This behavior can be explained by the evolution on the reprocessing of X-ray emission by the SN ejecta, as it becomes ionized by the X-ray radiation and bound-free absorption by oxygen is reduced. We estimate the time $t_{\rm ion}$ when the ionization rate of the ejecta via bound-free absoprtion exceeds the recombination rate. We adopt (i) an oxygen-poor SN ejecta of mass fraction $X_{\rm O}\sim 0.1$ expected from a low-mass helium star \citep[e.g.,][]{Dessart21}, (ii) a SN ejecta profile of $\rho_{\rm ej}\propto r^{-1}$ as in eq. (\ref{eq:rho_ej}), (iii) an ionizing photon spectrum of $\nu L_\nu=L_0(\nu/\nu_{\rm th})^{\alpha}$ with $\alpha<1$ and $h\nu_{\rm th}=0.87$ keV being the ionization threshold of oxygen K-shell electrons ($h$ is Planck constant), and (iv) an electron temperature of $\sim 10^6$ K expected for X-ray heated gas, with an associated recombination coefficient of $\alpha_{\rm rec}\approx 10^{-12}\ {\rm cm^3\ s^{-1}}$ \citep{Ho19,Margutti19}. The ionization and recombination rates are then
\begin{eqnarray}
    \mathcal{R}_{\rm ion}&\approx& \int_{\nu_{\rm th}}^{\infty}\frac{L_\nu}{h\nu} d\nu \approx \frac{L_0}{h\nu_{\rm th}(1-\alpha)}, \\
    \mathcal{R}_{\rm rec} &\approx& \int_0^{R_{\rm ej}}4\pi r^2 \alpha_{\rm rec} n_e n_{\rm O} dr \nonumber \\
    &\approx& \int_0^{R_{\rm ej}}\alpha_{\rm rec} \frac{4\pi r^2 X_{\rm O} \rho_{\rm ej}^2}{32m_p^2} dr \approx \frac{\alpha_{\rm rec}X_{\rm O}M_{\rm ej}^2}{32\pi m_p^2 R_{\rm ej}^3},
\end{eqnarray}
where $n_e\approx \rho_{\rm ej}/(2m_p), n_{\rm O}\approx X_{\rm O}\rho_{\rm ej}/(16m_p)$ are the number densities of electrons and oxygen nuclei. Equating these rates and using $R_{\rm ej}\approx t\sqrt{10E_{\rm ej}/3M_{\rm ej}}$, we find
\begin{eqnarray}
    t_{\rm ion}&\approx&\sqrt{\frac{3M_{\rm ej}}{10E_{\rm ej}}}\left[ \frac{\alpha_{\rm rec}X_{\rm O}M_{\rm ej}^2 h\nu_{\rm th}(1-\alpha)}{32\pi m_p^2 L_0} \right]^{1/3} \nonumber \\
    &\sim& 20\ {\rm day}\left(\frac{1-\alpha}{0.5}\right)^{\!\!1/3}\left(\frac{L_{\rm 0}}{10^{43}\ {\rm erg\ s^{-1}}}\right)^{\!\!-1/3} \left(\frac{X_{\rm O}}{0.1}\right)^{\!\!1/3} \nonumber \\
    &&\times \left(\frac{M_{\rm ej}}{0.5~M_\odot}\right)^{\!\!7/6}\left(\frac{E_{\rm ej}}{10^{51}\ {\rm erg}}\right)^{\!\!-1/2},
\end{eqnarray} 
which is roughly consistent with AT2018cow. In our model, the X-rays may originate from the non-thermal emission from the wind nebula and/or from the accreting central NS. We note that a 3.7$\sigma$ quasi-periodic feature in the soft X-ray light curve of 225 Hz was reported for AT 2018cow \citep[][but see also \citealt{Zhang22}]{Pasham21}. If true, this requires X-ray emission from a confined region ($\lesssim 10^{8}$ cm), and favors a significant contribution of X-rays from the accreting NS.

Radio observations of FBOTs indicate the existence of dense circumstellar matter (CSM) at radii of $10^{16}$--$10^{17}$ cm. While the inferred mass-loss rates are highly uncertain due to parameter degeneracies in radio afterglow modeling and the unknown CSM speed $v_{\rm CSM}$, they are inferred to be in the range $\sim 10^{-5}$--$10^{-3}~M_\odot\ {\rm yr}^{-1}(v_{\rm CSM}/10^3\ {\rm km\ s^{-1}})$ \citep{Ho19,Margutti19,Coppejans20,Ho20,Yao22}.

The CSM density profile of some FBOTs shows a robust steepening at radii $\sim 3\times 10^{16}$ cm \citep{Bright22}, which indicates that the dense CSM was created not too long before the explosion. The low-mass helium star progenitor invoked to explain the fast evolution can also naturally explain this density steepening. The envelope of low-mass helium stars can expand after core helium depletion from $\lesssim 1~R_\odot$ to up to $\sim 100~R_\odot$, triggering Case BC mass transfer onto the companion with peak rates of $\sim 10^{-4}~M_\odot\ {\rm yr}^{-1}$ at 0.1-1 kyrs before core-collapse \citep[][Wu \& Tsuna in prep]{tauris15_ultra_stripped_SNe, Wu22, Ercolino24}. Moreover, \citet{Wu22} show that, within decades from the explosion (after core Ne/O ignition), the rapid expansion of the outer envelope leads to extremely large mass-loss rates of $\gtrsim 10^{-2}~M_\odot\, \rm yr^{-1}$.
If mass transfer is not conservative, these processes can create dense CSM with a mass-loss rate that is a fraction of the mass-transfer rate. Depending on the ejection speed of the CSM, the density steepening at $\sim 3\times10^{16}\rm\,cm$ may be consistent with a delay time of $\sim 10 \mr{\,yrs}\, (v_{\rm CSM}/10^3{\rm \, km\,s^{-1}})^{-1}$. Typical equatorial mass loss from the binary's L2 point may have speeds of 10--100 km s$^{-1}$, whereas the typical speeds of the disk wind from super-Eddington accretion onto the companion star are in the range of 100-1000 km s$^{-1}$. Thus, the dense CSM might be consistent with either the L2 mass-loss in the core carbon burning phases, or the accretion disk wind launched during the extreme mass transfer in the core Ne/O burning phases.

\subsection{Multiple Encounters}
\label{sec:multiple_encounter}
Our light curve modeling only considers the main disruption event, where the entire star is disrupted and forms an accretion disk. However, the outcome of the encounter should be diverse, mainly dependent on the penetration factor $r_{\rm cl}/r_{\rm T}$. For larger distances of closest approach $r_{\rm cl}$ comparable to $R_*$, it is likely that the NS can have multiple passages through the star, partially disrupting the star at each passage before the star is fully disrupted \citep[e.g.,][]{Kremer22_TDE,Kremer23}. 

These multiple encounters are likely more relevant for events with longer $t_{\rm TDE}$, which we predict to have a brighter peak (Figures \ref{fig:rise_peak}, \ref{fig:rise_peak_BH}) as long as the ejecta is optically thick (see Sec. \ref{sec:light_curves}). When including the previous encounters, the energy injection will last longer by up to $t_{\rm TDE}$. Moreover, multiple encounters provides the opportunity for internal shocks between adjacent ejections to efficiently dissipate the kinetic energy of the disk wind. Hence, the rise of the light curve may instead be set by the energy injection history, and can be longer than what we have predicted assuming only the last disruption by up to $t_{\rm TDE}$. Furthermore, if the interval between each disruption is comparable or longer than the photon diffusion time through the SN ejecta (days to weeks), such events may display multiple bright peaks in the light curve, with each peak powered either by each accretion event or internal shocks generated by successive accretion-driven outflows.

In fact light curves of SLSNe are often found to have bumpy features \citep[e.g.,][]{Inserra17,Hosseinzadeh22} and/or pre-peak excess \citep[e.g.,][]{Nicholl16}, which are explained by variability of the central engine or additional circumstellar interaction. Our framework may naturally explain these features, as the typical duration of these bumps (10s of days) is similar to the diffusion timescale for a $3$--$10~M_\odot$ ejecta. A similar possibility was also raised by \cite{Hirai22}, although detailed modeling of the accretion power was not done. Combining our light curve modeling with the hydrodynamical simulations for the case of multiple encounters would be an important future work to test these hypotheses.

\subsection{Possible caveats and future avenues}

We focused on the case of disruptions following stripped envelope SNe of Type Ibc, and did not consider the case for Type II SNe. However, this channel is highly unlikely to happen for Type II SNe from supergiants, as it would require a large separation $\gtrsim 1000~R_\odot$ for the binary. The chance for encounter would hence be extremely rare. In case it happens, this may power Type II (super)luminous SNe without clear signatures of CSM interaction, or long-lasting Type II SNe that likely require a sustained heating source \citep[e.g.,][see also \citealt{Matsumoto24}]{Arcavi17}. 

We have estimated the energy injection considering the disk wind, but the energetics of the wind can depend on the assumed power-law index $p$, as the dynamical range of the disk $r_{\rm disk}/R_{\rm NS}\sim 10^4$--$10^5$ is very large. A plausible range suggested from simulations of $p=0.4$--$0.6$ \citep{Yuan12} changes the peak luminosity by about a factor of a few. We also ignored the possible contribution from matter that reaches the NS surface. As NSs have hard surfaces in contrast to BHs, the energy dissipated at the surface may contribute to the wind power, with an uncertain efficiency that can depend on the effects of neutrino cooling, and NS's magnetic fields and rotation. This excess energy may be added to the kinetic energy of the disk wind as it expands, and may further enhance the luminosity of the transient. Finally, if an asymmetric jet is launched from the accreting NS/BH, it can generate asymmetries in the energy injection and viewing-angle effects in the light curve that are not captured in this study \citep[e.g.,][]{Akashi20,Akashi21}.

For the companion, we adopted a mass-radius relation valid for massive stars in the main sequence. In reality the star would inflate due to interaction with the ejecta \citep{Wheeler75,Hirai18,Hober22}, which is not taken into account in this model. However this would not significantly alter the result, as we are interested in full disruptions that accrete the bulk of the star rather than the low-mass surface material subject to inflation. Nevertheless, envelope inflation would enhance ``grazing events" with $r_{\rm p}$ comparable to the (inflated) $R_*$, where the NS periodically accretes mass from the inflated companion at each orbit for a companion's surface thermal timescale. Such mechanism was considered for a recent stripped envelope SN 2022jli, that showed a 12.4-day periodic modulation for $\sim 200$ days after the light curve peak \citep[][see also \citealt{Moore23}]{Chen24}.

We suggest that the bulk of the companion will be tidally disrupted by the NS/BH by this encounter, but the detailed dynamics of the disruption process can be further affected by the stellar structure. If the companion has a well-developed core, the finite angular momentum of the core can trigger accretion of the core material onto the NS/BH. Such accretion phenomena have been proposed to potentially power energetic transients like (ultra-)long gamma-ray bursts \citep[e.g.,][]{Zhang01,Hutchinson-Smith24}, or even a stable X-ray bright remnant \citep{Everson24}.

In our model, the typical SLSN population with month-long rise and magnitudes of $\approx -21$ mag \citep{DeCia18,Gomez24} are better reproduced by the models with larger ejecta mass of $10~M_\odot$. While this may explain the preference for heavier progenitors and low-metallicity environments in SLSNe, the detailed effects of metallicity on the progenitor are beyond the scope of this work. The metallicity can also influence the binary evolution, as stronger stellar winds tend to expand the orbit and hence reduce the chance of NS-companion close encounter \citep[e.g.,][]{Renzo19}. Furthermore, unstable mass transfer leading to stellar merger may be less common for lower metallicities. This can be because (i) low-metallicity accretors more efficiently restoring thermal equilibrium after mass transfer instead of inflating \citep{demink08}, (ii) by suppression of efficient orbit shrinking caused by mass loss from the outer Lagrangian point \citep[][Appendix A]{Lu23}, or (iii) the envelope of lower metallicity donors having less-negative binding energy at the onset of Roche-lobe overflow which occurs at a later evolutionary stage \citep{Klencki20, Marchant21}. An in-depth work with binary population synthesis calculations will be needed to explore the detailed metallicity dependence of our model.

Finally, predictions for non-thermal counterpart emission would be important for distinguishing the models for (super-)luminous SNe and FBOTs. There are detections of radio counterparts in some SLSNe \citep{Eftekhari19,Margutti23}, as well as a gamma-ray counterpart in SN 2022jli \citep{Chen24} and a nearby SLSN 2017egm \citep{Li24}. In our model the shock formed between the disk wind and the SN ejecta is collisionless, and acceleration of particles to relativistic energies is expected. We plan to explore the non-thermal emission from our scenario in future work.

\section{Conclusion}
\label{sec:conclusion}
We have developed a model for transients arising from a newborn compact object (NS or BH) from a Type Ibc SN dynamically encountering a main-sequence companion. The presence of a main-sequence companion is common for stripped-envelope SNe, and such encounters occur when the NS/BH is kicked towards the companion with a velocity comparable or larger than the orbital velocity. We focused on the case where the companion is eventually disrupted by the NS/BH with a fraction of the star being bound to the NS/BH, inspired from recent hydrodynamical simulations. The super-Eddington accretion of bound material onto the NS/BH results in powerful outflows, which collide with and (re-)energize the SN ejecta.

We calculated the emission powered by this ``central engine" of an accreting NS/BH, by constructing a one-zone model that follows the thermodynamics of the SN ejecta as well as detailed efficiency of converting the dissipated outflow energy into radiation\footnote{The light curve source code is publicly available at \url{https://github.com/DTsuna/binaryTDE.git}.}. The transient becomes much brighter than normal Type Ibc SNe, with peak luminosities of the order of $\sim 10^{44}$ erg s$^{-1}$ and timescales of days to months. The optical luminosity and duration are consistent with what are observed in luminous Type Ibc SNe (except for the brightest and longest end of SLSNe), and FBOTs like AT2018cow.

We further carried out a Monte-Carlo analysis to estimate the fraction of main-sequence disruptions following a SN event, finding that the fraction is sensitive to the orbital parameters like binary separation and companion mass. For binary separations and companion masses expected from previous population synthesis calculations, we conclude that such disruptions can occur for $0.1$--$2$\% of stripped-envelope SNe, which is compatible with the rates of these luminous transients.

When compared to other existing models, our model also has a potential advantage that it can explain the peculiar properties observed in these transients, such as the late-time hydrogen line emission in SLSNe and FBOTs, as well as the bumpy features in the light curves of SLSNe. We however note that (i) there are existing suggestions in the framework of the magnetar model that can potentially address some of these challenges \citep[e.g.,][]{Kasen16,Moriya22,Zhu24,Gottlieb24}, and (ii) our prediction on bumpy features is still a proof-of-concept, which is yet to be verified in this work with a simple setup. Future works combining hydrodynamical simulations of such repeated tidal encounters with our emission model would be desired.

%\begin{acknowledgments}
\section*{Acknowledgements}
    We thank the anonymous referee for comments that greatly improved the manuscript. We also thank Iair Arcavi, Edo Berger, Jim Fuller, Daniel Kasen, Kazumi Kashiyama, Kyle Kremer, Raffaella Margutti, Selma de Mink, Matt Nicholl, Luc Dessart, Takashi Moriya, and Anthony Piro for valuable discussions. D. T. is supported by the Sherman Fairchild Postdoctoral Fellowship at Caltech. This research benefited from interactions that were funded by the Gordon and Betty Moore  Foundation through Grant GBMF5076. 
%\end{acknowledgments}

\appendix
\section{Radiative power of the shocked wind}
\label{sec:radiative power}
Here we describe our semi-analytical formulation for obtaining the radiative power from the shocked disk wind. The key is to understand which part of the disk wind, having a broad range of velocities, would dominate the radiative power. The velocity derivative of the wind mass-loss rate at $v=v_{\rm wind}$ is 
\begin{eqnarray}
    \left(\frac{d\dot{M}_{\rm wind}}{dv}\right)_{v=v_{\rm wind}} =\left(-\frac{d\dot{M}_{\rm in}}{dr}\frac{dr}{dv}\right)_{v=v_{\rm wind}}= (2p) \frac{|\dot{M}_{\rm disk}|}{v_{\rm wind}}\left(\frac{v_{\rm wind}}{v_{\rm disk}}\right)^{-2p} \propto v_{\rm wind}^{-2p-1},
\end{eqnarray}
where $v\propto r^{-1/2}$, $\dot{M}_{\rm in}\propto r^{p}\propto v^{-2p}$ is the mass inflow rate, and $\dot{M}_{\rm disk}=-M_{\rm disk}/t_{\rm visc}, v_{\rm disk}=\sqrt{GM_{\rm NS}/r_{\rm disk}}$ are respectively the accretion rate and wind velocity at the characteristic disk radii $r_{\rm disk}$. Thus for $0<p<1$, the wind mass ($\int dv (d\dot{M}_{\rm wind}/dv)$) is mainly carried by the slowest (outermost) part of the wind, whereas the wind kinetic energy ($\int dv (d\dot{M}_{\rm wind}/dv)(v^2/2)$) is carried by the fastest (innermost) part near the NS. Cooling of shock-heated gas occurs via interaction between the heated electrons and ions/photons, whose rate increases with the density of the wind and is hence lower for faster winds. Thus which part of the wind dominates the radiative output is non-trivial, and we therefore estimate this taking into account the efficiency of gas cooling as described below.

The relevant timescales are the dynamical timescale of the shocked wind
\begin{eqnarray}
    t_{\rm dyn} \approx \frac{R_{\rm neb}}{dR_{\rm neb}/dt},
\end{eqnarray}
and the cooling timescale $t_{\rm cool}$ set by the harmonic mean of the two cooling times defined by that of inverse Compton (IC) scattering of soft SN photons and free-free emission, which are both dependent on the upstream wind velocity
\begin{eqnarray}
    t_{\rm cool}(v_{\rm wind}) = \left[\frac{1}{t_{\rm cool, IC}(v_{\rm wind})} + \frac{1}{t_{\rm cool, ff}(v_{\rm wind})}\right]^{-1}.
\end{eqnarray}
The radiation conversion efficiency is then defined as
\begin{eqnarray}
    \epsilon_{{\rm rad},v}(v_{\rm wind}) = \frac{1}{1+t_{\rm cool}/t_{\rm dyn}}, \label{eq:eps_rad}
\end{eqnarray}
and the radiative power of the shocked wind is
\begin{eqnarray}
    L_{\rm rad, wind} = \int_{v_{\rm crit}}^{v_{\rm max}} \epsilon_{{\rm rad},v}\frac{dL_{\rm wind}}{dv}dv = \int_{v_{\rm crit}}^{v_{\rm max}} \epsilon_{{\rm rad},v}\left(\frac{v^2}{2}\frac{d\dot{M}_{\rm wind}}{dv}\right)dv,
\end{eqnarray}
where $v_{\rm crit}=\sqrt{GM_{\rm NS}/r_{\rm crit}}$, and $v_{\rm max}=\sqrt{GM_{\rm NS}/R_{\rm NS}}$. The kinetic power of the shocked wind $L_{\rm wind}$ is in eq. (\ref{eq:Lwind}), and the \textit{global} radiation conversion efficiency that goes into eq. (\ref{eq:dEraddt}) is then
\begin{equation}\label{eq:wind_radiative_efficiency_global}
    \epsilon_{\rm rad}=L_{\rm rad, wind}/L_{\rm wind}.
\end{equation}
Since gas cooling is faster for denser wind/ejecta with higher optical depth, $\epsilon_{\rm rad}$ is initially close to unity and drops below unity at later times (see also Figure \ref{fig:Mej_comp}). The time when $\epsilon_{\rm rad}$ starts to drop below unity mainly depends on $M_{\rm ej}$. For our ejecta models of $M_{\rm ej}=1,3,10~M_\odot$, the ranges of epochs when $\epsilon_{\rm rad}$ drops to $0.9~(0.5)$ are $40$--$120$ ($50$--$130$), $70$--$160$ ($90$--$180$), and $100$--$280$ ($150$--$330$) days from SN respectively.

\subsection{Timescales of Gas Cooling}
Here we obtain the cooling timescales of the shocked gas $t_{\rm cool, ff}(v_{\rm wind}), t_{\rm cool ,IC}(v_{\rm wind})$. We assume that the ions and electrons in the shock-heated gas reach equipartition due to either Coulomb relaxation or plasma instabilities driven by the (collisionless) shock \citep[e.g.,][]{Spitkovsky:2008aa, Kato:2010aa, Park15, Crumley:2019aa}, with temperature and density given by shock compression (with adiabatic index 5/3) as
\begin{eqnarray}
    T_{\rm w, sh}(v_{\rm wind}) &\approx& \frac{3}{16}\frac{\mu m_p v_{\rm wind}^2}{k_B} \sim 1.4\times 10^9\ {\rm K}\left(\frac{\mu}{0.62}\right)\left(\frac{v_{\rm wind}}{10^9\ {\rm cm\ s^{-1}}}\right)^2 \label{eq:Tw_sh}\\
    \rho_{\rm w, sh}(v_{\rm wind}) &\approx& 4\times \frac{(v\times d\dot{M}_{\rm wind}/dv)|_{v=v_{\rm wind}}t_{\rm dyn}}{4\pi R_{\rm neb}^3} = 4\times \frac{(2p)\dot{M}_{\rm disk}(v_{\rm wind}/v_{\rm disk})^{-2p}t_{\rm dyn}}{4\pi R_{\rm neb}^3},
\end{eqnarray}
where $k_B$ is the Boltzmann constant, and the density is derived by dividing the swept-up mass of the wind having velocity $v_{\rm wind}$, which is $\approx (v\times d\dot{M}_{\rm wind}/dv)|_{v=v_{\rm wind}}t_{\rm dyn}$, by the (compressed) volume $4\pi R_{\rm neb}^2(R_{\rm neb}/4)$. As a typical case we adopt a fully ionized gas of solar abundance (hydrogen mass fraction $X_{\rm H}=0.7$), with a mean molecular weight $\mu=0.62$.

The free-free cooling timescale is
\begin{eqnarray}
    t_{\rm cool,ff}(v_{\rm wind}) &=& \frac{3\rho_{\rm w, sh}k_BT_{\rm w, sh}/(2\mu m_p)}{\Lambda(T_{\rm w, sh}) (X_{\rm H}\rho_{\rm w, sh}/m_p)^2} \propto \rho_{\rm w, sh}^{-1}T_{\rm w, sh}^{1/2} \propto v_{\rm wind}^{1+2p} \propto v_{\rm wind}^2\ (p=0.5),
\end{eqnarray}
where we adopt a free-free cooling function $\Lambda(T)\approx 1\times 10^{-23}\ {\rm erg\ s^{-1}\ cm^3}~(T/10^7~{\rm K})^{1/2}$ \citep{Sutherland93}.

We estimate the IC cooling timescale by scattering with external cold photons in the SN ejecta. The diffusion time in the hydrogen-rich shocked wind, with width $\Delta R_{\rm neb}(<R_{\rm neb})$, is 
\begin{eqnarray}
   t_{\rm d, neb}&\sim&\frac{\kappa_{\rm wind} M_{\rm w,sh}}{4\pi R_{\rm neb}^2} \times \frac{\Delta R_{\rm neb}}{c} \sim 2\ {\rm day}\left(\frac{\Delta R_{\rm neb}}{R_{\rm neb}}\right)\left(\frac{\kappa_{\rm wind}}{0.3\ {\rm cm^2\ g^{-1}}}\right)\left(\frac{M_{\rm w,sh}}{0.1~M_\odot}\right)\left(\frac{R_{\rm neb}}{10^{15}\ {\rm cm}}\right)^{-1}.
\end{eqnarray}
Given that $\Delta R_{\rm neb}/R_{\rm neb}< 1$, $t_{\rm d, neb}$ is shorter than the diffusion time in the (unshocked) ejecta (eq. \ref{eq:tdiff_ej}) 
\begin{eqnarray}
   t_{\rm diff}&\sim& 6\ {\rm day}\left(\frac{\kappa}{0.07\ {\rm cm^2\ g^{-1}}}\right)\left(\frac{M_{\rm ej}}{3~M_\odot}\right)\left(\frac{R_{\rm ej}}{2\times 10^{15}\ {\rm cm}}\right)^{-1},
\end{eqnarray}
or the dynamical timescale $t_{\rm dyn}$. Hence, we expect to first order that the shocked wind and ejecta would share the same radiation density set by the radiation content of the ejecta $u_{\rm rad}\approx E_{\rm rad}/(4\pi R_{\rm ej}^3/3)$, where $E_{\rm rad}$ and $R_{\rm ej}$ are both solved by the one-zone modeling in Section \ref{sec:onezone_model}.

If the radiation energy density $u_{\rm rad}$ does not significantly change as the shocked gas cools, the timescale is given as
\begin{eqnarray}
    \tilde{t}_{\rm cool, IC}(v_{\rm wind})\approx \frac{3\mu_e m_e c}{8\mu u_{\rm rad}\sigma_{\rm T}} \left[1+\frac{4k_B T_{\rm w, sh}}{m_e c^2}\right]^{-1} \sim 0.49\ {\rm yr}\left(\frac{\mu_e}{\mu}\right)\left(\frac{u_{\rm rad}}{\rm erg\ cm^{-3}}\right)^{-1}\left[1+\frac{4k_B T_{\rm w, sh}}{m_e c^2}\right]^{-1}, \label{eq:t_cool_IC}
\end{eqnarray}
where $m_e$ is the electron mass, $\mu_e=2/(1+X_{\rm H})$, and $\sigma_{\rm T}$ is the Thomson cross section. 

Whether the approximation of constant $u_{\rm rad}$ holds is set by the Compton $y$-parameter, which characterizes the change in photon energy density as they diffuse through the shocked gas. The $y$-parameter is evaluated as the product of the mean number of scatterings $N_{\rm scat}$ and the fractional energy gain per scattering \citep{Rybicki79}
\begin{eqnarray}
    y &\approx& N_{\rm scat}\times \frac{4}{3}\langle \beta^2\gamma^2\rangle \approx \tau_{\rm scat}(1+\tau_{\rm scat})\times \frac{4}{3}\langle \beta^2\gamma^2\rangle, 
\end{eqnarray}
where $\tau_{\rm scat}$ is the scattering optical depth of the (hydrogen-rich) shocked wind
\begin{eqnarray}
    \tau_{\rm scat} &=& \frac{\kappa_{\rm scat}t_{\rm dyn}}{4\pi R^2_{\rm neb}} \int_{v_{\rm crit}}^{v_{\rm max}} dv {d \dot{M}_{\rm wind}\over d v},
\end{eqnarray}
with $\kappa_{\rm scat}=0.2(1+X_{\rm H})$ cm$^2$ g$^{-1}$ being the scattering opacity, $\beta$ is the electron thermal velocity scaled by speed of light, and $\gamma=1/\sqrt{1-\beta^2}$ is the Lorentz factor. With $v_{\rm wind}$ ranging from $v_{\rm crit}\approx$ (a few -- $10)\times 10^8$ cm s$^{-1}$ (see Figure \ref{fig:nebula_prop}) to $v_{\rm max}\approx 1.2\times 10^{10}$ cm s$^{-1}$, the corresponding thermal energy of electrons derived from eq. (\ref{eq:Tw_sh}) spans from non-relativistic ($\sim 10$--$100$ keV) to mildly relativistic ($\sim 20$ MeV) energies. In the non-relativsitic limit ($\beta\ll 1$ and $\gamma\approx 1$), the typical $\beta\gamma$ of the electrons scales as $\beta\gamma \propto \sqrt{k_B T_{\rm w, sh}/m_e c^2}\propto v_{\rm wind}$, while in the relativistic limit ($\beta\approx 1$ and $\gamma\gg 1$) it scales as $\beta\gamma \propto k_BT_{\rm w, sh}/m_e c^2\propto v_{\rm wind}^2$. For a wind mass-loss profile $(d\dot{M}_{\rm wind}/dv)dv\propto v^{-2p-1}dv$, we can crudely approximate the $\beta\gamma$ distribution of the shock-heated electrons as a piecewise power law
\begin{eqnarray}
    f(\beta\gamma)d(\beta\gamma) \sim 
    \begin{cases}
       A(\beta\gamma)^{-2p-1}d(\beta\gamma) & ({\rm for}\ (\beta\gamma)_{\rm min}< \beta\gamma < 1) \\
       A(\beta\gamma)^{-p-1}d(\beta\gamma) & ({\rm for}\ 1\leq \beta\gamma < (\beta\gamma)_{\rm max}),
    \end{cases}
\end{eqnarray}
with $A$ being the normalization constant
\begin{eqnarray}
    A = \left[\frac{1}{2p}\left((\beta\gamma)_{\rm min}^{-2p}-1\right) + \frac{1}{p}\left(1-(\beta\gamma)_{\rm max}^{-p}\right) \right]^{-1}, \label{eq:app_A}
\end{eqnarray}
which for $(\beta\gamma)_{\rm min}\ll 1 \ll (\beta\gamma)_{\rm max}$ is $\approx 2p(\beta\gamma)_{\rm min}^{2p}$, or simply $\approx (\beta\gamma)_{\rm min}$ for $p=0.5$.

On the other hand, the fractional energy gain per scattering $4\langle\beta^2\gamma^2\rangle/3$ is evaluated as
\begin{eqnarray}
    {4\left<\beta^2\gamma^2\right>\over 3} &=& \frac{4}{3}\int_{(\beta\gamma)_{\rm min}}^{(\beta\gamma)_{\rm max}} (\beta\gamma)^2f(\beta\gamma) d(\beta\gamma) = \frac{4A}{3} \left[\frac{1}{2-2p} \left(1-(\beta\gamma)_{\rm min}^{2-2p}\right)  + \frac{1}{2-p} \left((\beta\gamma)_{\rm max}^{2-p}-1\right) \right], \label{eq:app_energy_gain}
\end{eqnarray}
which for $(\beta\gamma)_{\rm min}\ll 1 \ll (\beta\gamma)_{\rm max}$ is $\approx [8p/3(2-p)](\beta\gamma)_{\rm min}^{2p}(\beta\gamma)_{\rm max}^{2-p}$, or $\approx (8/9)(\beta\gamma)_{\rm min}(\beta\gamma)_{\rm max}^{1.5}$ for $p=0.5$. It is important to note here that the energy gain is dominantly contributed by the hottest electrons with $\beta\gamma\approx (\beta\gamma)_{\rm max}$.

The values $(\beta\gamma)_{\rm min}, (\beta\gamma)_{\rm max}$ are set from the temperatures $T_{\rm w, sh}(v_{\rm crit})$, $T_{\rm w, sh}(v_{\rm max})$ as
\begin{eqnarray}
    (\beta\gamma)_{\rm min} &=& \sqrt{3}\times \sqrt{k_BT_{\rm w, sh}(v_{\rm crit})/m_ec^2} \\
    (\beta\gamma)_{\rm max}&=&\sqrt{12}\times k_BT_{\rm w, sh}(v_{\rm max})/m_ec^2,
\end{eqnarray}
where the prefactors come from the root mean square of $\beta\gamma$ in the nonrelativistic and relativistic regime \citep{Rybicki79}. Using these pre-factors recovers the well-known formula in the single-temperature case when $f(\beta\gamma)$ is a delta function, of $4\left<\beta^2\gamma^2\right>/3=4k_BT_{\rm w, sh}/m_ec^2 (=16(k_BT_{\rm w, sh}/m_ec^2)^2)$ in the non-relativistic (relativistic) limit.

When $y\gtrsim 1$, the energy density $u_{\rm rad}$ is significantly modified by Comptonization exponentially with $y$, invalidating the timescale in eq. (\ref{eq:t_cool_IC}) that assumes constant $u_{\rm rad}$. We therefore include an exponential dependence on $y$ in the cooling time and formulate $t_{\rm cool, IC}$ as 
\begin{eqnarray}
t_{\rm cool, IC} =  \frac{1}{{\rm min}\left[e^y, \frac{T_{\rm w,sh}(v_{\rm max})}{T_{\rm ej}}\right]}  \tilde{t}_{\rm cool, IC},    
\end{eqnarray}
which smoothly connects to the limit of $y\ll 1$ where $t_{\rm cool, IC}=\tilde{t}_{\rm cool, IC}$. The timescale is capped by the temperature ratio $T_{\rm w,sh}(v_{\rm max})/T_{\rm ej}$ where $T_{\rm ej}=(u_{\rm rad}/a)^{1/4}$ is the photon temperature in the ejecta, as once photons are upscattered to energies comparable to $3k_BT_{\rm w, sh}(v_{\rm max})$, energy exchange by IC scattering would be inefficient.

\bibliography{references}
\end{document}